\documentclass[%
reprint,
superscriptaddress,
amsmath,amssymb,
floatfix,
]{revtex4-2}

\usepackage{graphicx}
\usepackage{dcolumn}
\usepackage{bm}
\usepackage{physics}
\usepackage{soul}
\usepackage{times}
\usepackage[breaklinks,colorlinks,urlcolor=blue,citecolor=blue,linkcolor=blue]{hyperref}
\usepackage[mathlines]{lineno}
\usepackage{amsfonts,amssymb,amsmath,mathbbol}              
\usepackage{comment}   

\setcitestyle{super}
\begin{document}

\title{Chern dartboard insulator: sub-Brillouin zone topology and skyrmion multipoles} 
 
\author{Yun-Chung Chen }
\affiliation{Department of Physics and Center for Theoretical Physics, National Taiwan University, Taipei 10607, Taiwan }

\author{Yu-Ping Lin}
\affiliation{Department of Physics, University of California, Berkeley, California 94720, USA}
 
\author{Ying-Jer Kao}
\email{yjkao@phys.ntu.edu.tw}
\affiliation{Department of Physics and Center for Theoretical Physics, National Taiwan University, Taipei 10607, Taiwan }
\affiliation{Center for Quantum Science and Technology,  National Taiwan University, Taipei 10607, Taiwan }
\affiliation{National Center for High-Performance Computing, Hsinchu City 30076, Taiwan}

\begin{abstract}
Topology plays a crucial role in many physical systems, leading to interesting states at the surface. 
The paradigmatic example is the Chern number defined in the Brillouin zone that leads to the robust gapless edge states.
%
%
%
%
%
%
%
%
Here we introduce the reduced Chern number, defined in subregions of the Brillouin zone (BZ), and construct a family of Chern dartboard insulators (CDIs) with quantized reduced Chern numbers in the sub-BZ (sBZ) but with trivial bulk topology.
CDIs are protected by mirror symmetries and exhibit distinct pseudospin textures, including (anti)skyrmions, inside the sBZ. 
These CDIs host exotic gapless edge states, such as M\"{o}bius fermions and midgap corner states, and can be realized in photonic crystals. 
Our work opens up new possibilities for exploring sBZ topology and nontrivial surface responses in topological systems.
\end{abstract}

\maketitle

\vspace{-2ex} 

\section*{Introduction} \label{sec1}

Chern insulators are classic examples of non-interacting systems with nontrivial bulk topology~\cite{Haldane1988},  in which the quantized Hall conductivity observed in transport experiments is related to the first Chern number defined in the Brillouin zone (BZ)~\cite{Qi2006}. 
The associated quantum anomalous Hall effect has also been observed ~\cite{Chang2013,Liu2016,He2018,Deng2020,Serlin2020} along with robust gapless edge states~\cite{Liu2016,He2018,Ozawa2019,Wang2009}. 
In this work, we show that the nontrivial topology can appear locally in the BZ, even when the global topology is trivial. 
The idea relies on the concept of sub-Brillouin zone (sBZ) topology where the topological invariant is defined in a fraction of the BZ.
We introduce a family of delicate topological systems~\cite{Nelson2021,Nelson2022}, termed as the \textit{Chern dartboard insulators} (CDIs), which exhibit the nontrivial sBZ topology. 
The $n$-th order Chern dartboard insulators, CDI$_n$ for short, have quantized first Chern numbers inside $1/2n$ of the BZ. These \textit{reduced Chern numbers} are protected by $n$ mirror symmetries (Fig.~\ref{summary}a).
These systems cannot be captured by the theories of tenfold way~\cite{Kitaev2009,Ryu2010}, symmetric indicators~\cite{Kruthoff2017,Po2017,Po2020} or topological quantum chemistry~\cite{Bradlyn2017,Cano2018}. 
In addition, all the CDIs exhibit multicellular and even noncompact topology~\cite{Nelson2021,Nelson2022,Schindler2021}, i.e., the Wannier functions cannot be entirely localized to $\delta$-functions due to the reduced Chern numbers.
Similar to the returning Thouless pump (RTP) insulators, all the CDIs can be captured by the quantized Berry phases along the high-symmetry lines (HSLs), and the topology can be trivialized by adding trivial atomic bands into either the occupied or unoccupied space.

Interestingly, for the two-band CDIs, skyrmion multipoles appear as a manifestation of the sBZ topology in the BZs. Contrary to Chern insulators that exhibit meron-antimeron pairs, the BZs of CDIs consist of HSLs which pin the pseudospins in the same direction. The (anti)skyrmions live in the sBZs bounded by the HSLs, thereby exhibiting the multipole structures (Fig.~\ref{pseudospin}). The total number of (anti)skyrmions in the sBZ indeed corresponds to the reduced Chern number. 

Here we observe that with the sBZ topology, all the CDIs host gapless edge states, even with M\"{o}bius fermions and midgap corner states in certain cases. 
However, contrary to Chern insulators, by including weak disorders (compared to the bulk gap) that obey mirror symmetries, one can in general gap out the edge states. 
In this sense, the edge (corner) states are as robust as the two-dimensional (2D) weak TIs or the multipole-moment insulators protected by crystalline symmetries~\cite{Benalcazar2017,Benalcazar2017_2,Benalcazar2019,Schindler2019}. 
Nevertheless, under the sharp boundary condition, the gapless edge states can be protected at certain high-symmetry edges.

\section*{Analyses and Results}

\textbf{Theory.} We aim at finding the possible topology protected by $n$ mirror symmetries $\mathcal{M}_1,\mathcal{M}_2,...,\mathcal{M}_n$. These symmetries divide the BZ into the irreducible BZs, which become the fundamental domain to define the topology. A possible realization is to consider the systems with the same mirror symmetry representation $\mathcal{M}_1,...,\mathcal{M}_n=\sigma_z\otimes I$,
\begin{eqnarray} \label{equation mirror}
\mathcal{M}_iH(\bm{k})\mathcal{M}_i^{-1}=H(R_i\bm{k}),
\end{eqnarray}
where $\sigma_z$ and $I$ are the Pauli and identity matrix, and $R_i$ represent the mirror reflections in the $\bm{k}$ space. Here, the basis orbitals are chosen such that the mirror symmetry representation is diagonal. Therefore, the projection matrix onto the occupied space at the HSLs is in a block diagonal form. Notice that the system also has $C_n$ symmetry with trivial representation $C_n=I$.

Next, we consider the models in which all the occupied states at the HSLs have the same mirror representations.
The blocks in the projection matrix are thus composed of zero and identity matrices, which correspond to the unoccupied and occupied space, respectively. 
Up to a $\bm{k}$-dependent $U(1)$ phase, the HSLs are mapped to a point in the Hilbert space, and each sBZ enclosed by the HSLs is topologically equivalent to a compact manifold. 
The first Chern number is thus well-defined in the irreducible BZ,
\begin{equation} \label{equation stokes}
\mathcal{C}_n=-\frac{1}{2\pi}\int_\text{irBZ} d^2k \Tr F_{xy},
\end{equation}
where $F_{xy}$ is the non-Abelian Berry curvature, and $\mathcal{C}_n$ is the reduced Chern number of the $n$-th order CDIs. 

The simplest CDI$_1$ is protected by one mirror symmetry and has the quantized reduced Chern number inside half of the BZ. 
There are two types of CDI$_1$: Type I  has the opposite mirror representations at the two different HSLs.
An example  is given by the two-band tight-binding Hamiltonian with the mirror symmetry $\mathcal{M}_y=\sigma_z$,
\begin{eqnarray} \label{equation n=1 typeI}
H^I_{1}(\bm{k})&&=\cos k_x\sin k_y\sigma_x+\sin k_x\sin k_y\sigma_y \nonumber\\
&&+(m+\cos k_y)\sigma_z,
\end{eqnarray}
where $m$ is a tunable parameter. In this model, the two HSLs sit at $k_y=0,\pi$.
The basis orbitals consist of an $s$ orbital and a $p_y$ orbital. 
At $-1<m<1$, this model has a quantized reduced Chern number $\mathcal{C}_1=1$ inside the upper half BZ, and has a flat-band limit at $m=0$.
Type I  CDI$_1$  has a quantized bulk polarization if the total number of the occupied bands is  odd~\cite{Benalcazar2017}. 

\begin{figure}[t]
\begin{center}
\includegraphics[width=\columnwidth]{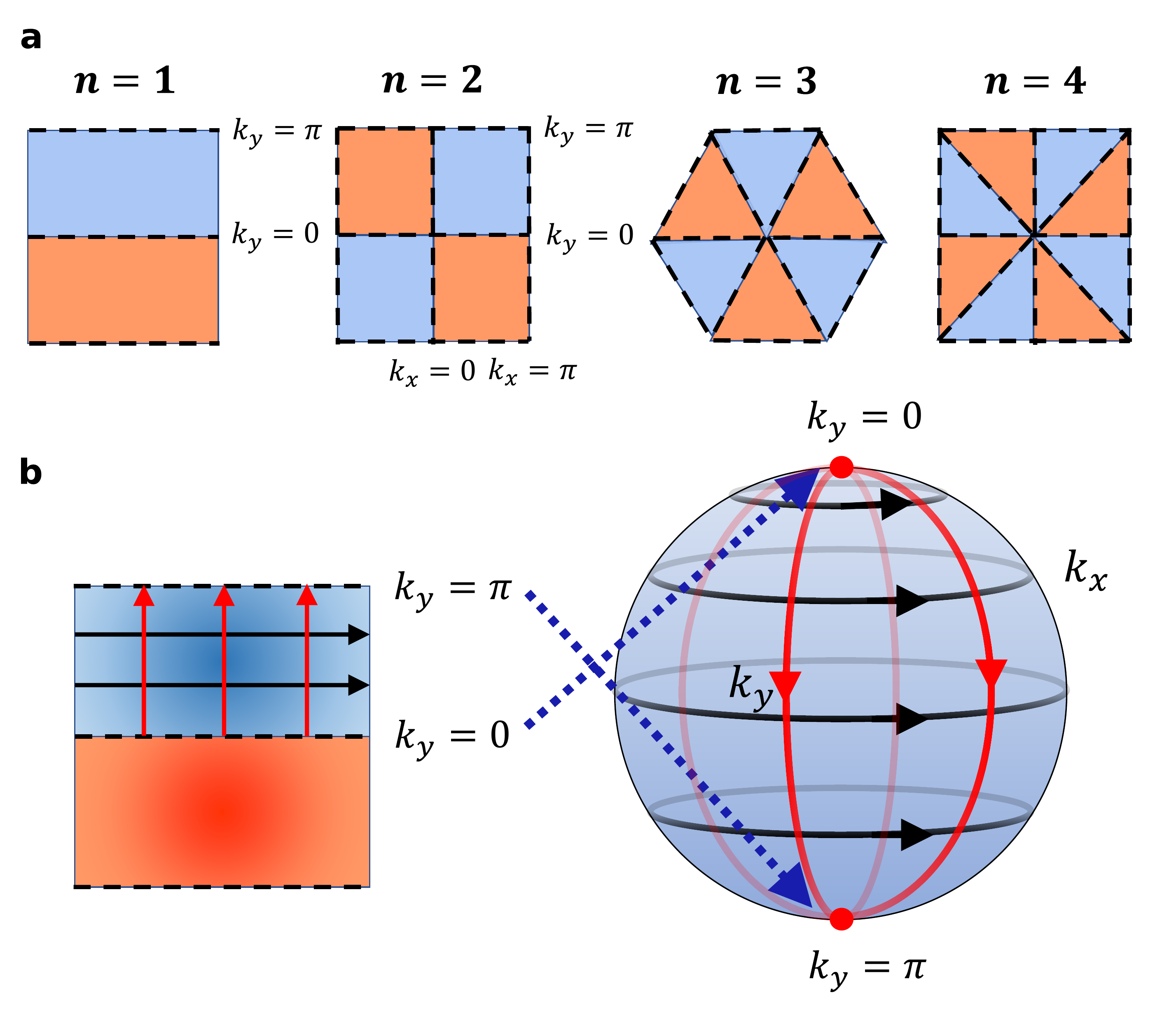}
\caption[]{\textbf{Illustration of Chern dartboard insulators.} $\textbf{\textsf{a}}$ Chern dartboard insulators with different orders.  The black dashed lines denote the high-symmetry lines of mirror symmetries in the Brillouin zone. The Chern number is quantized inside the regions enclosed by the high-symmetry lines. Here, the regions with blue and red color have reduced Chern number $\mathcal{C}_n=1,-1$. $\textbf{\textsf{b}}$ Mercater projection of Type I $n=1$ Chern dartboard insulator.}
\label{summary}
\end{center}
\end{figure}

\begin{figure}[t]
\begin{center}
\includegraphics[width=\columnwidth]{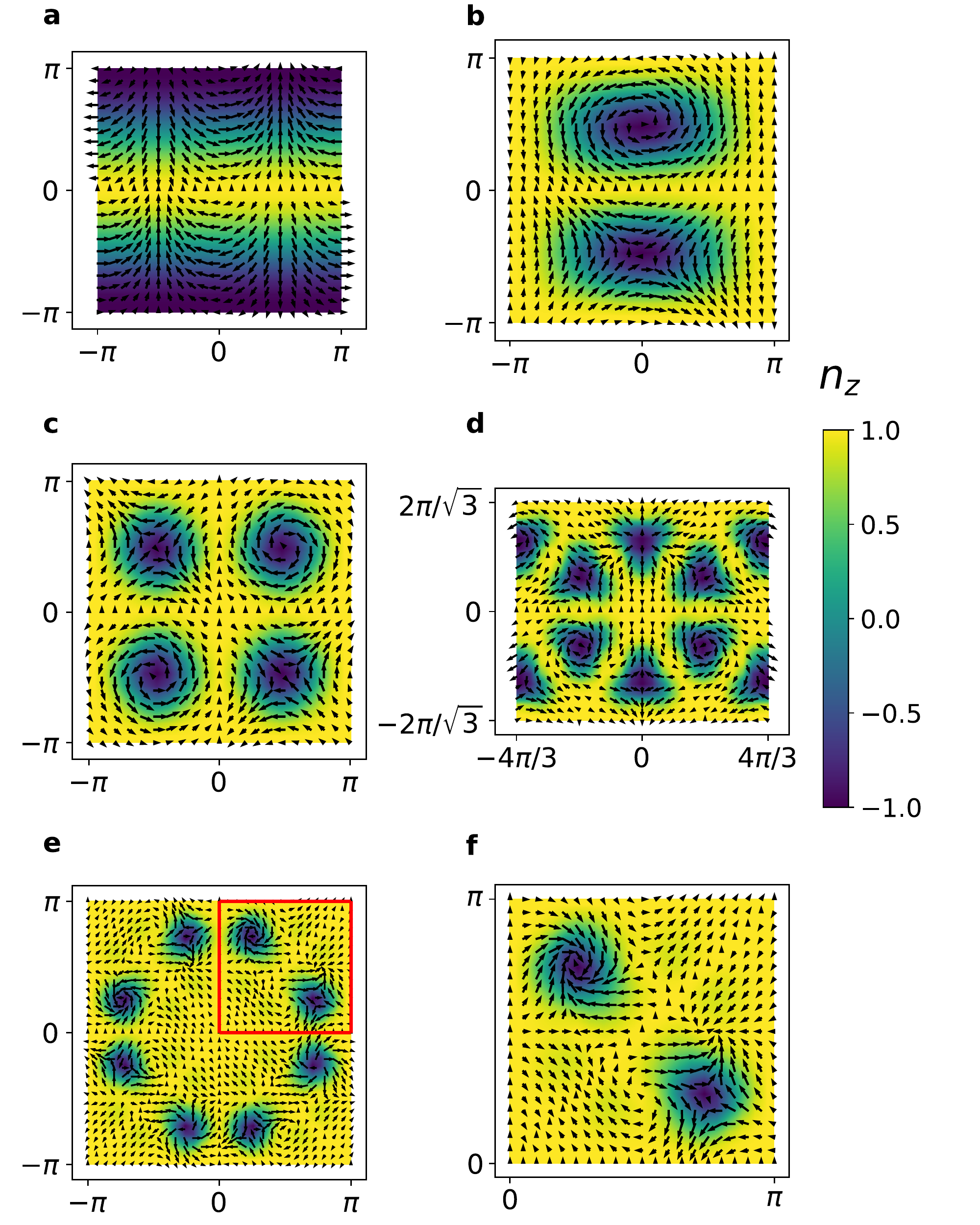}
\caption[]{\textbf{Pseudospin textures of the Chern dartboard insulators.} The vector field represents  the components of $n_x$ and $n_y$, and the color represents the $n_z$ component. $\textbf{\textsf{a}}$ Type I $n=1$ Chern dartboard insulator in the flat-band limit. The reduced Chern number $\mathcal{C}_1=1$ arising from the winding around the south pole at $k_y=\pi$. $\textbf{\textsf{b}}$ Type II $n=1$ Chern dartboard insulator in the flat-band limit. The reduced Chern number $\mathcal{C}_1=1$ arising from the skyrmion living inside half of the BZ. $\textbf{\textsf{c}}$ $n=2$ Chern dartboard insulator. $\textbf{\textsf{d}}$ $n=3$ Chern dartboard insulator. $\textbf{\textsf{e}}$ $n=4$ Chern dartboard insulator. $\textbf{\textsf{f}}$ The zoom-in subplot for the red frame in $\textbf{\textsf{e}}$.}
\label{pseudospin}
\end{center}
\end{figure}

On the other hand, type II CDI$_1$ has the same mirror representations at the two different HSLs. 
A flat-band model is given by
\begin{eqnarray}
H^{II}_1(\bm{k})&&=\frac{1}{2}(1+\cos k_x)\sin 2k_y\sigma_x+\sin k_x\sin k_y\sigma_y \nonumber\\
&&+\frac{1}{2}[(1+\cos k_x)(\cos 2k_y-1)+2]\sigma_z.
\label{TypeII}
\end{eqnarray}
Interestingly, the bulk polarization $P_x=\int^{2\pi}_0 \Tr A_xdk_x$ shows the RTP behavior~\cite{Nelson2021,Nelson2022} in both cases.
The RTP invariant is given by the difference of polarizations along the HSLs $\Delta P_x=P_x(k_y=\pi)-P_x(k_y=0)$. Notably, this invariant can be related to the quantized reduced Chern number through the Stokes theorem.

The sBZ topology can be visualized by mapping (\ref{equation n=1 typeI})  to the Bloch sphere with  $k_x\rightarrow\phi,k_y\rightarrow\theta$~\cite{Alexandradinata2022,Zhu2022}. 
This mapping is only meaningful when $k_y\in[0,\pi]$.
The upper half BZ is mapped to the entire sphere through the Mercator projection, thereby hosting the reduced Chern number $\mathcal{C}_1=1$ (Figs.~\ref{summary}b and~\ref{pseudospin}a). 
Meanwhile, the lower half BZ $k_y\in[-\pi,0]$ is also mapped to the entire sphere, but now with a negative sign $\mathcal{C}_1=-1$ due to mirror symmetry. 
The quantization of the reduced Chern number inside the half BZ can be understood as follows: under the mirror symmetry $\mathcal{M}_y=\sigma_z$, the $k_y=0$ and $\pi$ HSLs are mapped to the north and south poles, respectively, as they have opposite representations. 
Therefore, the half BZ has a topology equivalent to $S^2$, and the reduced Chern number $\mathcal{C}_1$ is well-defined (Fig.~\ref{summary}b).
In contrast to Type I CDI$_1$, Type II CDI$_1$ has well-defined skyrmions inside the half BZ, as shown in Figure~\ref{pseudospin}b. The sBZ topology is more complicated as both the HSLs are mapped to the north pole, and it can be directly related to the existence of skyrmions.

Higher-order CDIs, on the other hand, are quite different from the  CDI$_1$s as they cannot be captured by the RTP invariant.
In particular, the symmetry representations of the valence bands (or the conduction bands) at all the HSLs are exactly the same as the symmetry representation of one of the basis orbitals.
Figures~\ref{pseudospin}c,~\ref{pseudospin}d and~\ref{pseudospin}e plot the peudospin textures of the two-band higher-order CDIs.
All these cases have blue-centered skyrmions or antiskyrmions inside the irreducible BZs, which lead to the nontrivial reduced Chern number $\mathcal{C}_n=1$. 
Finally, we note that there are two types of CDI$_3$s with different configurations of the irreducible BZs. 
Here we plot Type I CDI$_3$ that corresponds to the configuration in Fig.~\ref{summary}a. 

By inspecting the symmetry representations at the HSLs alone, one cannot detect the CDIs.
However, the CDIs still have different band representations from the ones with $\delta$-like Wannier functions, since the homotopic inequivalence occurs from the quantized reduced Chern number. 
Moreover, the $n=2,4,6$ CDIs are  noncompact atomic insulators, where the orthonormal Wannier functions cannot be strictly local and compact~\cite{Schindler2021}. 
Note that these models are quite different from the cases studied in Ref.~\cite{Schindler2021}, where the noncompactness arises from  the obstruction of the lattices that leads to obstructed atomic insulators. 
Here, the CDIs are not obstructed and the noncompactness arises due to the multicellularity.

\textbf{Gapless edge states.} Similar to the Chern insulators, the appearance of gapless edge states in CDIs can be explained by the domain walls, with the exception of Type I CDI$_1$s. The domain walls arise because of the band inversions inside the irreducible BZs. For Type I CDI$_1$s, there are no isolated band-inversion points and the gapless edge states only exist in the directions perpendicular to the HSLs. 
For other CDIs, there must exist isolated band-inversion points inside the irreducible BZ, owing to the quantized reduced Chern number $\mathcal{C}_n$. 
Inside the irreducible BZ, the topology shares similar behavior with regular Chern insulators. 
If one tries to close the gap by creating massive Dirac cones at the band-inversion points, the minimal low-energy Dirac Hamiltonian with mirror symmetry representation $\sigma_z\otimes I$ can be expressed as
\begin{eqnarray} \label{equation dirac}
H(\bm{k},r)=k_x\Gamma_x+k_y\Gamma_y+m(r)\sigma_z\otimes I,
\end{eqnarray}
where the three matrices $\Gamma_x,\Gamma_y,\sigma_z\otimes I$ anticommute with each other, and $m(r)$ is a position-dependent mass term. 
Since  $m(r)<0$  inside the bulk, the gapless edge states appear as the domain walls between the bulk and the vacuum in which $m(r)>0$~\cite{Hasan2010}.
In the two-band models, the band-inversion points are exactly associated with the skyrmion centers in Fig.~\ref{pseudospin}.

\begin{figure}
\begin{center}
\includegraphics[width=\columnwidth]{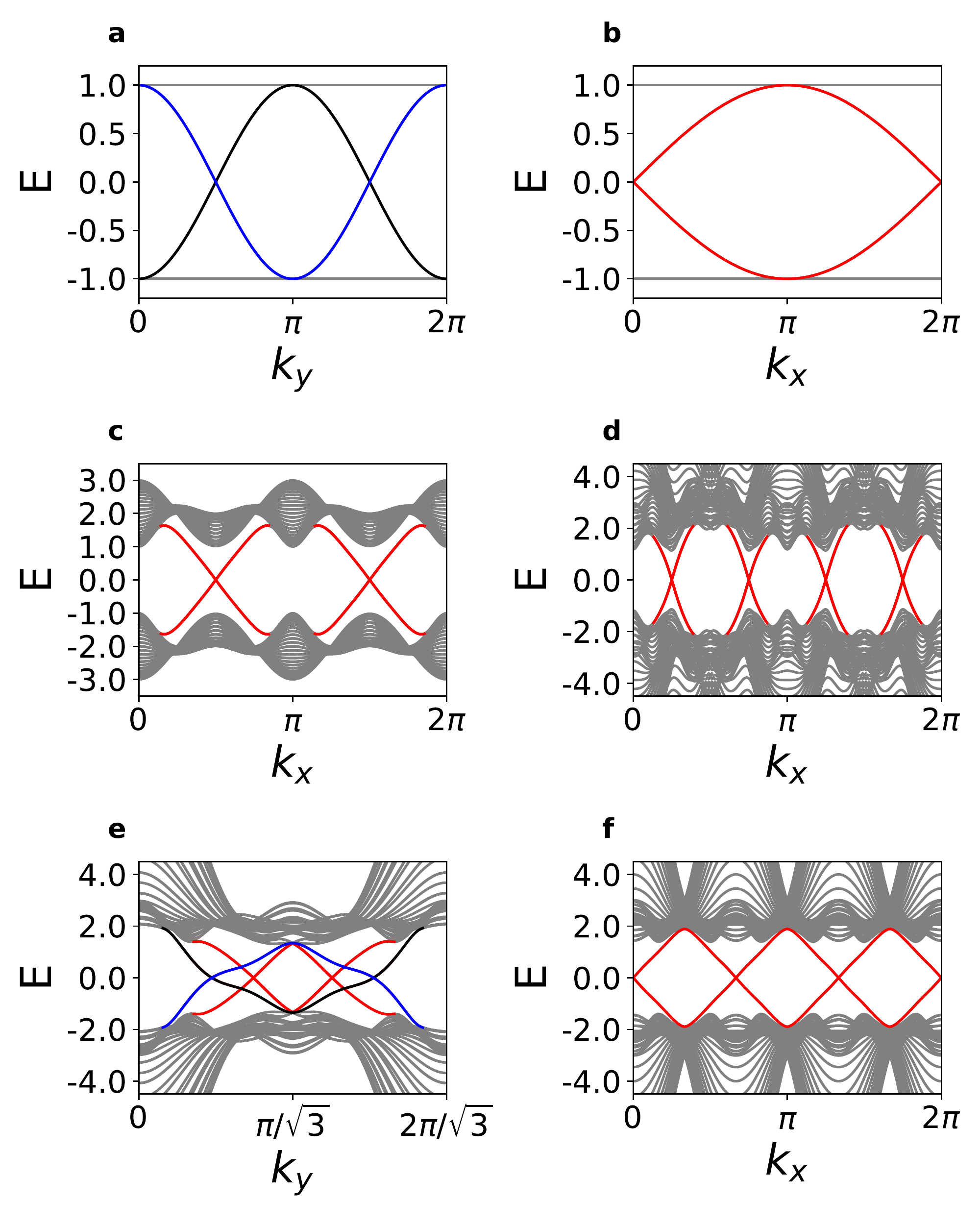}
\caption[]{\textbf{Edge states of Chern dartboard insulators}. The nanoribbon band structures are shown in all subfigures. The doubly degenerate edge states localized separately at the two opposite edges are highlighted by red color. $\textbf{\textsf{a}}$ Type I and II CDI$_1$s  with edges along the $y$-direction in the flat-band limit. The edge states localized at the left (right) edge are denoted by blue (black) color. $\textbf{\textsf{b}}$ Type II CDI$_1$  with edges along the $x$-direction in the flat-band limit.  $\textbf{\textsf{c}}$ CDI$_2$  with edges along the $x$-direction. $\textbf{\textsf{d}}$ CDI$_4$ with edges along the $x$-direction. $\textbf{\textsf{e}}$ CDI$_3$ with edges along the $y$-direction, which correspond to the zigzag edges. The edge states localized at the left (right) edge are denoted by blue (black) color. $\textbf{\textsf{f}}$ CDI$_3$  with edges along the $x$-direction, which correspond to the flat edges.}
\label{edge}
\end{center}
\end{figure}

It is worth emphasizing that there is a fundamental difference between the Chern insulators and the CDIs. In a Chern insulator with $\mathcal{C}=1$, we have only one band-inversion point inside the BZ. 
Thus, the chiral gapless edge states appear due to the domain walls between the bulk and the vacuum.
However, for CDIs with $\mathcal{C}_n=1$, we have $2n$ band-inversion points with opposite signs of reduced Chern number inside the BZ under mirror symmetries. 
It follows that the edges host $n$ gapless edge states with positive chirality and $n$ gapless edge states with negative chirality near $E=0$, regardless of the edge terminations. 
The bulk-boundary correspondence for all the CDIs, except for Type I CDI$_1$, can be written as,
\begin{eqnarray} \label{equation bbc}
N^+_e=N^-_e=n\mathcal{C}_n,
\end{eqnarray}
where $N^\pm_e$ is the minimal number of gapless edge states with positive (negative) chirality near $E=0$.
Here, we emphasize again that this picture only works in the low-energy region. One can expect that the gapless edge states with positive and negative chiralities may rejoin at high energy such that they are disconnected from the bulk bands. 
In this sense, the gapless edge states in the CDIs are as robust as the delicate TIs.
In Fig.~\ref{edge}, we plot the nanoribbon band structures for all the CDIs. It is easy to observe that the bulk-boundary correspondence Eq.~(\ref{equation bbc}) is satisfied. Finally, the edge states in Fig.~\ref{edge}c, Fig.~\ref{edge}d and Fig.~\ref{edge}f are doubly degenerate because of mirror symmetry that relates the two boundaries.

\begin{figure}
\begin{center}
\includegraphics[width=\columnwidth]{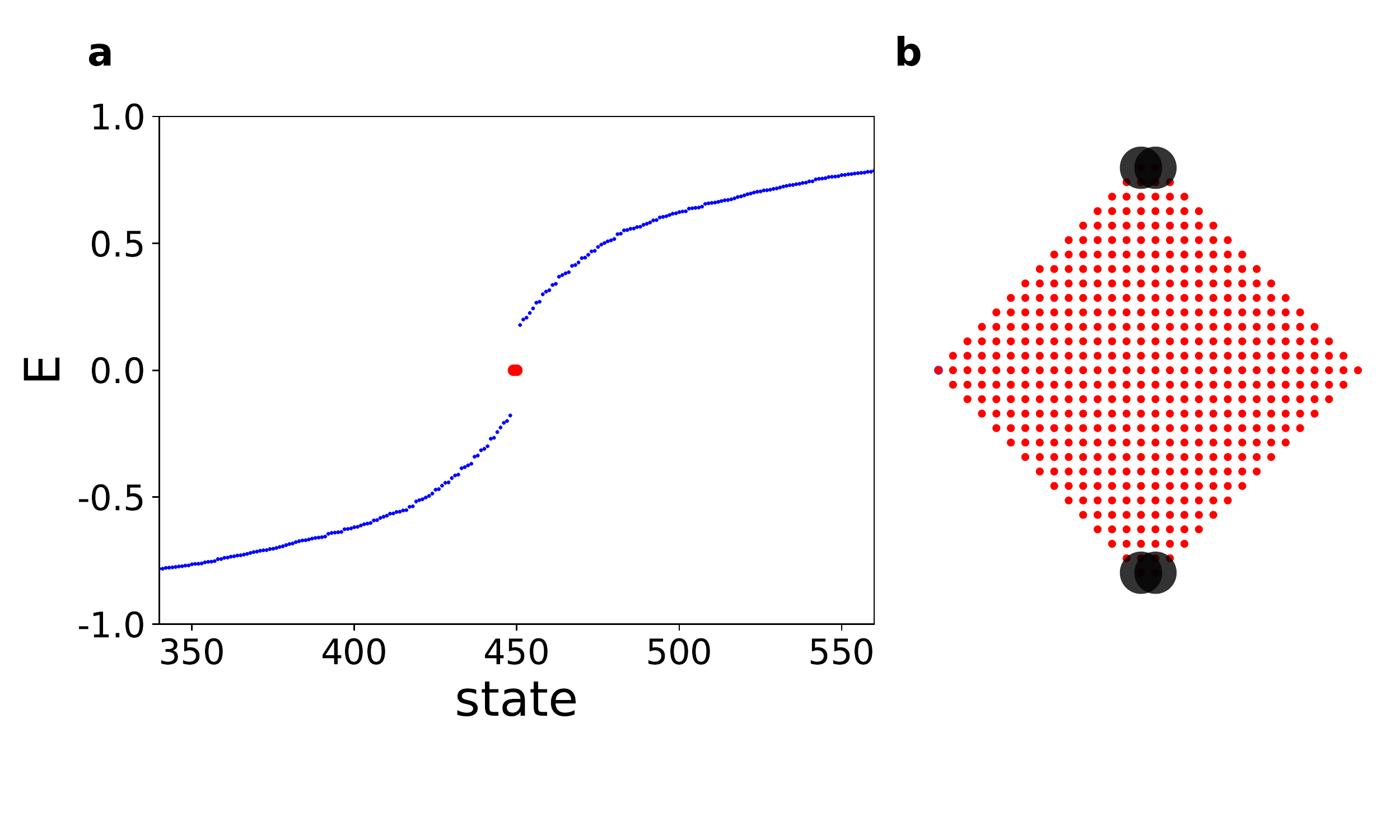}
\caption[]{\textbf{Midgap corner states in Type I $\bm{n=1}$ Chern dartboard insulator.} $\textbf{\textsf{a}}$ The nanoflake energy spectrum. $\textbf{\textsf{b}}$ The probability distribution of the two corner states. In the flat-band limit, the left and right corners of the nanoflake also host midgap corner states, but they are not robust. To explicitly show it, we set $m=0.2$ and add the nearest-neighbor coupling $0.3\cos k_x\sigma_z$ to Eq. (\ref{equation n=1 typeI}).}
\label{corner}
\end{center}
\end{figure}

\textbf{Midgap Corner States.} By cutting the edges in the (1,1) and (1,-1) directions of Type I CDI$_1$, we obtain two protected midgap corner states with quantized $1/2$ charges. 
Notice that the unit cell is preserved, and the upper and lower corners have two atoms instead of one atom. 
Contrary to the 2D weak TIs where there are plenties of midgap edge states, Type I CDI$_1$ host two protected midgap corner states at the upper and lower corners (Fig.~\ref{corner}). 
This is similar to the midgap corner states in the model with a quantized polarization~\cite{Ezawa2018}, where the corner states appear at the corners that correspond to the direction of polarization. 
Therefore, Type I CDI$_1$ also serves as the simplest two-band models with protected midgap corner states.

\textbf{M\"{o}bius Fermions.} Interestingly, the edge states  parallel to the $x$-axis in Type II CDI$_1$ show even more striking phenomenon. 
We expect that there are two chiral gapless edge states with opposite chirality near $E=0$, regardless of the edge terminations due to the band inversions inside the irreducible BZ. 
In the flat-band limit, the edge Hamiltonian describes the phase transition point of the four-band SSH model, and the edge states form the M\"{o}bius fermions~\cite{Shiozaki2015} (Fig. \ref{edge}b). 
The energy of the two edge states are $E_1(k_x)=\sin (k_x/2)$ and $E_2(k_x)=-\sin (k_x/2)$. 
Clearly, they show the M\"{o}bius twist: the energies are only periodic in $k_x\in[0,4\pi]$, and a single edge Dirac cone appears at $k_x=0$. Since the edge Dirac cone is protected by the quantized reduced Chern number, the Dirac cone cannot disappear although the position can shift from $k_x=0$. 
The $e^{ik_x/2}$ dependence is also clearly shown in the edge states (see Supplementary Note 2 for details). 
In addition to Type II CDI$_1$, the M\"{o}bius fermions also appear in both types of CDI$_3$s (Fig.~\ref{edge}f). 
Finally, we note that the M\"{o}bius fermions can only appear with odd numbers of Dirac cones in the CDIs.

\section*{Discussions}
We introduce a novel concept of sBZ topology that manifests itself in different classes of CDIs. 
%
%
CDIs host gapless edge states in general and can even develop nontrivial M\"{o}bius fermions or midgap corner states in certain cases.
Although here we only consider  specific examples of the sBZ topology,  one can easily generalize the same argument to systems in higher dimensions or with different symmetries/constraints. 
The sBZ topology thus opens a fertile area for new topological systems with nontrivial surface responses.
The different physical properties from the global topology make the realization and classification of the sBZ topology a new frontier of research in topological materials.
Thanks to the recent advances in  detecting local Berry curvature in various systems~\cite{Wimmer2017,Schuler2020},  and  in realizing topological phases in nanophotonic silicon ring resonators~\cite{Leykam2020}, 
the realization and observation of CDIs and their exotic edge states is expected in the near future.

\section*{Methods}

\small

\textbf{Conventions and definitions.} With translational symmetry, the second quantized tight-binding Hamiltonian can be written into the Bloch form,
\begin{eqnarray}
\hat{H}=\sum_{\bm{k}}c^{\dagger}_{i,\bm{k}}\big[H(\bm{k})\big]_{ij}c_{j,\bm{k}},
\end{eqnarray}
where 
\begin{eqnarray}
c_{j,\bm{k}}=\frac{1}{\sqrt{N_t}}\sum_{\bm{R}}e^{i\bm{k}\cdot \bm{R}}c_{j,\bm{R}}
\end{eqnarray}
is the electron annihilation operator. Here, $j=1,...,2N$ labels the basis orbitals and spins, $\bm{R}$ labels the unit cell position, and $N_t$ is the total number of the unit cells. We use the convention that the Bloch Hamiltonian $H(\bm{k})$ is periodic under a translation of a reciprocal lattice vector $\bm{G}$:
\begin{eqnarray}
H(\bm{k})=H(\bm{k}+\bm{G}).
\end{eqnarray}
The intra-cell eigenstates are defined by:
\begin{eqnarray}
H(\bm{k})|u^l(\bm{k})\rangle=E^l(\bm{k})|u^l(\bm{k})\rangle,
\end{eqnarray}
where $l=1,...,2N$ are the band indices and $E^l(\bm{k})$ is the eigenenergy.
Note that one can generically choose a smooth and periodic gauge for the eigenstates of CDIs since the total Chern number is zero. Considering the half-filling band insulators, the intra-cell states can be decomposed into the valence states $|u^l_v\rangle$ and the conduction states $|u^l_c\rangle$, where $l=1,...,N$. The Bloch state is given by $|\psi^l(\bm{k})\rangle=e^{i\bm{k}\cdot\bm{R}}|u^l(\bm{k})\rangle$.

The non-Abelian Berry connection for the valence bands is defined as:
\begin{eqnarray}
\bm{A}_{lm}(\bm{k})=i\langle u^l_v(\bm{k})|\bm{\nabla}|u^m_v(\bm{k})\rangle,
\end{eqnarray}
and the non-Abelian Berry curvature in two dimensions is:
\begin{eqnarray}
F_{xy,lm}=\partial_xA_{y,lm}-\partial_yA_{x,lm}-i\big[A_x,A_y\big]_{lm}.
\end{eqnarray}

\textbf{Skyrmion number.} The topology of two-band CDIs can be visualized using the psudospin textures. We first expand the Bloch Hamiltonian into the Pauli matrices:
\begin{eqnarray} \label{equation expansion}
H(\bm{k})=\sum_id_i(\bm{k})\sigma_i,
\end{eqnarray}
where $\sigma_i$ with $i=x,y,z$ are the Pauli matrices. The reduced Chern number can be defined as the degree of the map from the irreducible BZ to $S^2$:

\begin{eqnarray} \label{equation chern}
\mathcal{C}_n=-\frac{1}{4\pi}\int_\text{irBZ} d^2k\bm{n}\cdot\left(\frac{\partial\bm{n}}{\partial k_x}\times\frac{\partial\bm{n}}{\partial k_y}\right)
\end{eqnarray}
where $\bm{n}(\bm{k})=\bm{d}(\bm{k})/|\bm{d}(\bm{k})|$ are the unit vectors that define the space of $S^2$. The reduced Chern number measures how many times the irreducible BZ wraps around $S^2$.

The peudospin textures of CDIs are plotted in Fig.~\ref{pseudospin}.  The first Chern number can be calculated as the sum of the indices around either $\bm{n}_0=(0,0,1)$ or $\bm{n}_0=(0,0,-1)$ inside the BZ, with a sign difference~\cite{Gero2021}. Note that this choice is just for convenience as one can easily do an arbitrary unitary transformation. The reduced Chern number is
\begin{eqnarray} \label{equation skyrmion}
\mathcal{C}_n=\sum_iS_i,
\end{eqnarray}
where $S_i$ is the index around the south pole $\bm{n}_0=(0,0,-1)$ with the blue center in the plot, and $i$ denotes the different skyrmions inside the BZ. Since the irreducible BZ is a two-dimensional manifold, the indices can be calculated simply as the winding numbers of the vectors around the south pole. Notice that the Poincaré–Hopf theorem constrains the total indices including those around the north pole $\bm{n}_0=(0,0,1)$ are summed to zero. This is consistent with Eq.~(\ref{equation chern}) since a nonzero Chern number implies the north pole and the south pole must be wrapped around nontrivially.

\textbf{Tight-binding models.} Here we list the two-band spinless tight-binding models used to construct the figures in the main text for CDIs. The $n=1,2,4$ CDIs are built in the simple square lattice with primitive lattice vectors $\bm{a}_1=(1,0)$ and $\bm{a}_2=(0,1)$ in the unit of a lattice constant. The CDI$_3$ is built in the triangular lattice with primitive lattice vectors $\bm{a}_1=(1,0)$ and $\bm{a}_2=(1/2,\sqrt{3}/2)$ in the unit of a lattice constant. The basis orbitals are all placed on the atoms. The numerical results of the midgap corner states and the nanoribbon band structures are performed using the P\textsubscript{YTH}TB package \cite{pythTB}.

Type I CDI$_1$:
\begin{eqnarray}
H^I_{1}(\bm{k})&&=\cos k_x\sin k_y\sigma_x+\sin k_x\sin k_y\sigma_y \nonumber\\
&&+(m+\cos k_y)\sigma_z,
\end{eqnarray}
where $m$ is a parameter. The basis orbital consists of a $s$ orbital and a $p_y$ orbital. The Hamiltonian has the mirror symmetry $\mathcal{M}_y=\sigma_z$. For $-1<m<1$, this model has a quantized reduced Chern number $\mathcal{C}_1=1$ inside the upper half BZ, see Fig.~\ref{summary}. The system has a flat-band limit when $m=0$. 

Type II CDI$_1$:
\begin{eqnarray}
H^{II}_1(\bm{k})&&=\cos k_x\sin 2k_y\sigma_x+\sin k_x\sin k_y\sigma_y \nonumber\\
&&+(m+\cos 2k_y-\cos k_x)\sigma_z,
\end{eqnarray}
where $m$ is a parameter. The basis orbital consists of a $s$ orbital and a $p_y$ orbital. The Hamiltonian has the mirror symmetry $\mathcal{M}_y=\sigma_z$. For $0<m<2$, this model has a quantized reduced Chern number $\mathcal{C}_1=1$ inside the upper half BZ, see Fig.~\ref{summary}. The system has a flat-band limit in the following form:
\begin{eqnarray}
H^{IIc}_1(\bm{k})&&=\frac{1}{2}(1+\cos k_x)\sin 2k_y\sigma_x+\sin k_x\sin k_y\sigma_y \nonumber\\
&&+\frac{1}{2}[(1+\cos k_x)(\cos 2k_y-1)+2]\sigma_z.
\end{eqnarray}

CDI$_2$:
\begin{eqnarray}
H_2(\bm{k})=&&-\sin k_x\sin 2k_y\sigma_x+\sin 2k_x\sin k_y\sigma_y \nonumber\\
&&+(m+\cos 2k_x+\cos 2k_y)\sigma_z.
\end{eqnarray}
where $m$ is a parameter. The basis orbital consists of a $s$ orbital and a $d_{xy}$ orbital. The Hamiltonian has two mirror symmetries $\mathcal{M}_x=\mathcal{M}_y=\sigma_z$. For $0<m<2$, this model has a quantized reduced Chern number $\mathcal{C}_2=1$ inside the upper right quarter of the BZ, see Fig.~\ref{summary}. We set $m=1.0$ for all the figures.

CDI$_4$:
\begin{eqnarray}
H_4(\bm{k})=(&&-\sin k_x\sin 4k_y+\sin 4k_x\sin k_y)\sigma_x \nonumber\\
+(&&\sin 2k_x\sin 4k_y-\sin 4k_x\sin 2k_y)\sigma_y \nonumber\\
+\big[&&m+\cos 2k_x+\cos 2k_y+\cos 4k_x \nonumber\\
&&+\cos 4k_y+4\cos k_x\cos k_y\big]\sigma_z,
\end{eqnarray}
where $m$ is a parameter. The basis orbital consists of a $s$ orbital and a second orbital that is odd under four mirror symmetries. The Hamiltonian has four mirror symmetries $\mathcal{M}_x=\mathcal{M}_y=\mathcal{M}_{x+y}=\mathcal{M}_{x-y}=\sigma_z$. For $2<m<4$, this model has a quantized reduced Chern number $\mathcal{C}_4=1$ inside the irreducible BZ, see Fig.~\ref{summary}. We set $m=3.0$ for all the figures.

Type I CDI$_3$:
\begin{eqnarray}
H_3(\bm{k})=\big[&&\sin \frac{5}{2}k_x\sin \frac{\sqrt{3}}{2}k_y-\sin 2k_x\sin \sqrt{3}k_y \nonumber\\
&&+\sin \frac{1}{2}k_x\sin \frac{3\sqrt{3}}{2}k_y\big]\sigma_x \nonumber\\
+\big[&&-\cos \frac{5}{2}k_x\sin \frac{\sqrt{3}}{2}k_y-\cos 2k_x\sin \sqrt{3}k_y \nonumber\\
&&+\cos \frac{1}{2}k_x\sin \frac{3\sqrt{3}}{2}k_y\big]\sigma_y \nonumber\\
+\bigg[&&m+\sum_{a=1}^6\big(e^{i\bm{t}_2(a)\cdot\bm{k}}+\frac{1}{2}e^{2i\bm{t}_2(a)\cdot\bm{k}}\big)\bigg]\sigma_z,
\end{eqnarray}
where $m$ is a parameter and $\bm{t}_2(a)=\sqrt{3}[\cos(\pi a/3-\pi/6),\sin(\pi a/3-\pi/6)]^T$. The basis orbitals consist of a $s$ orbital and a $f_{y(3x^2-y^2)}$ orbital. The Hamiltonian has three mirror symmetries $\mathcal{M}_y=C_3\mathcal{M}_yC_3^{-1}=C_3^2\mathcal{M}_yC_3^{-2}=\sigma_z$. The $\sigma_z$ term contains the hoppings with range $\bm{a}_1+\bm{a}_2$, $2(\bm{a}_1+\bm{a}_2)$, and also the ones generated by all the $C_6$ rotations. The $\sigma_x$ and $\sigma_y$ terms contain the hoppings with range $\bm{a}_1+2\bm{a}_2$ and the ones generated by three mirror symmetries. For $0<m<4.5$, this model has a quantized reduced Chern number $\mathcal{C}_3=1$ inside the irreducible BZ, see Fig.~\ref{summary}. We set $m=2.0$ for all the figures.

\section*{Data availability}

\scriptsize

The data for all the figures are all available by requesting.

\section*{Acknowledgments}

The authors thank Yi-Chun Hung for helpful discussions. Y.C.C. and Y.J.K. were partially supported by the Ministry of Science and Technology (MOST) of Taiwan under grants No. 108-2112-M-002-020-MY3, 110-2112-M-002-034-MY3, 111-2119-M-007-009, and by the National Taiwan University under Grant No. NTU-CC-111L894601. Y.P.L. received the fellowship support from the Emergent Phenomena in Quantum Systems (EPiQS) program of the Gordon and Betty Moore Foundation.

\section*{Author contributions}

Y.C.C conceived the ideas and performed the theoretical and numerical analysis. Y.P.L provided the idea to look at the pseudospin textures and M\"{o}bius fermions. Y.J.K. supervised the project. Y.C.C, Y.P.L and Y.J.K. wrote the manuscript.

\section*{Competing interests}

The authors declare no competing interests.

\end{document}


\title{\Large \textbf{Supplementary information for ``Chern dartboard insulator: sBZ topology and skyrmion multipoles"}} 
 
\date{}
 
\maketitle

\tableofcontents

\addcontentsline{toc}{section}{Supplementary Note 1. Mathematical Formulation of Chern Dartboard Insulators}
\section*{\centering Supplementary Note 1. Mathematical Formulation of Chern Dartboard Insulators} \label{sec1}

Consider $n$ mirror symmetries $\mathcal{M}_1,\mathcal{M}_2,...,\mathcal{M}_n$ with the same mirror symmetry representation $\mathcal{M}_1,...,\mathcal{M}_n=\sigma_z\otimes I$:
\begin{eqnarray} \label{equation mirror}
\mathcal{M}_iH(\bm{k})\mathcal{M}_i^{-1}=H(R_i\bm{k}),
\end{eqnarray}
where $R_i$ represent mirror reflections in $\bm{k}$ space. Using the Stokes theorem:
\begin{equation} \label{equation stokes}
-\frac{1}{2\pi}\oint\Tr\bm{A}\cdot d\bm{k}=-\frac{1}{2\pi}\int_\text{irBZ} d^2k \Tr F_{xy}=\mathcal{C}_n,
\end{equation}
where $\bm{A}$ and $F_{xy}$ are the non-Abelian Berry connection and curvature, and $\mathcal{C}_n$ is the reduced Chern number of the $n$-th order CDIs. The loop integral encloses the irreducible BZ. On the HSLs, the valence states are eigenstates of the Hamiltonian due to Eq.~(\ref{equation mirror}). Next, we consider the models which all the occupied states at the high-symmetry lines (HSLs) have the same mirror representations. As a result, the valence states are spanned by the basis orbitals:
\begin{equation} \label{equation basis}
|u^i_v(\bm{k}\in\text{HSLs})\rangle=\sum^N_{j=1}U_{ij}(\bm{k}\in\text{HSLs})|b^j\rangle,
\end{equation}
where $|b^j\rangle$ with $j=1,...,N$ are the basis orbitals that have the same mirror symmetry representation with the occupied states $|u^j_v(\bm{k}\in\text{HSLs})\rangle$. Here, the basis orbitals are chosen into the basis such that the mirror symmetry representation is diagonal. Since the total Chern number is zero, $U_{ij}(\bm{k})$ can be chosen to a smooth and periodic $U(N)$ gauge transformation. The key point is that the rank of the occupied states is equal to the rank of the basis states $|b^j\rangle$ such that $U_{ij}(\bm{k})$ is a well-defined $U(N)$ gauge transformation. By plugging Eq.~(\ref{equation basis}) into Eq.~(\ref{equation stokes}), it follows that,
\begin{equation} \label{equation $n=1$ CDI}
\frac{i}{2\pi}\int \Tr[U^{\dagger}_{\pi}\bm{\nabla}U_{\pi}-U^{\dagger}_0\bm{\nabla}U_0]\cdot d\bm{k}=\mathcal{C}_1,
\end{equation}
where $U_{\pi}$ and $U_0$ denote the gauge transformation at the two different HSLs, and
\begin{equation} \label{equation higher-order}
-\frac{i}{2\pi}\oint \Tr U^{\dagger}\bm{\nabla}U\cdot d\bm{k}=\mathcal{C}_n,
\end{equation}
for $n>1$. The reduced Chern number is thus quantized to integers. Note that the $\bm{k}$-dependent $U(1)$ phase is crucial to obtain the nonzero reduced Chern number.

\addcontentsline{toc}{section}{Supplementary Note 2. $n=1$ Chern Dartboard Insulator in the Compact Limit}
\section*{\centering Supplementary Note 2. $n=1$ Chern Dartboard Insulator in the Compact Limit} \label{sec2}

In this supplementary note, we discuss the analytical results of the CDI$_1$s in the compact limit. Here, the compact limit means that the model has the compact orthonormal Wannier functions with finite lattice sites. Note that for general parameters of the Hamiltonian, the Wannier functions decay exponentially rather than compactly localize.For the trivial atomic insulator, the compact limit is attained when the electrons entirely localize at each atom and the associated Wannier functions are $\delta$-functions. However, for the CDIs with multicellularity, it is impossible to adiabatically deform the models into the trivial atomic insulators. Nevertheless, it is possible for CDI$_1$s to attain the compact limit where the Wannier functions are strictly zero outside a finite domain. 

In the compact limit, the system has the following properties: (i) the correlation length is strictly zero, and the correlation function has a finite range; (ii) the excited spectrum is entirely flat since the excited degrees of freedom can be localized into a finite region. Therefore, there can't be dispersion of $\bm{k}$. The first property can be proved by expanding the correlation function into the basis of the orthonormal compact Wannier functions:
\begin{equation}
\begin{split}
C^{\bm{\alpha\beta}}_{ij}=\langle c^{\dagger}_{\bm{\alpha} i}c_{\bm{\beta} j}\rangle=\sum_{n\in\text{occ.},\bm{R}}\langle\bm{\alpha}_i| W^n_{\bm{R}}\rangle\langle W^n_{\bm{R}}|\bm{\beta}_j\rangle,
\end{split}
\end{equation}
where $\bm{\alpha,\beta}$ label the positions of two atomic lattice sites, and $i,j$ label the basis orbitals or spins. $W^n_{\bm{R}}$ are the Wannier functions constructed from the occupied space with position $\bm{R}$. Since the Wannier functions are compact, the correlation function must also be compact and become zero outside a finite domain. 

\textbf{Type I CDI\bm{$_1$}.} The Type I CDI$_1$ in the compact limit with $\mathcal{C}_1=1$ has the form:
\begin{equation}
\begin{split}
H^I_{1}(\bm{k})&=\cos k_x\sin k_y\sigma_x+\sin k_x\sin k_y\sigma_y +\cos k_y\sigma_z.
\end{split}
\end{equation}
Interestingly, this model can be simply related to the Bloch sphere through the mapping defined by the polar coordinate $k_x\rightarrow\phi,k_y\rightarrow\theta$. These two polar coordinates define the direction of the vector $\bm{n}$. It is clear that the upper half BZ is mapped to the entire sphere through the Mercator projection and therefore has reduced Chern number $\mathcal{C}_1=1$, see Fig.~2 in the main text. In the compact limit, the energy spectrum is entirely flat:
\begin{eqnarray}
E(\bm{k})=\pm|\bm{d}(\bm{k})|=\pm 1,
\end{eqnarray}
and the normalized valence state is given by:
\begin{eqnarray} \label{$n=1$ CDI state}
|u^I_v(\bm{k})\rangle=e^{-ik_y/2}\left(\begin{matrix}-e^{-ik_x}\sin\frac{k_y}{2}\\\cos\frac{k_y}{2}\end{matrix}\right).
\end{eqnarray}

Notice that a global periodic and smooth gauge is possible for the valence state as the total Chern number is zero under mirror symmetry. Since the normalization is a constant, we can easily obtain the orthonormal compact Wannier function at the origin $\bm{R}=0$:
\begin{eqnarray} 
|W_{\bm{R}=\bm{0}}\rangle=\frac{1}{(2\pi)^2}\int_{BZ}d^2ke^{i\bm{k}\cdot\bm{r}}|u^I_v(\bm{k})\rangle,
\end{eqnarray}
The Wannier function occupies four atomic lattice sites, see Supplementary Fig.~\ref{compact}a. The Wannier center is localized at Wyckoff position $b$, which shares the same representation with the 2D weak TI protected by mirror symmetry. One should note that these two models cannot be adiabatically deformed into each other due to the quantized reduced Chern number. 

\begin{figure}
\begin{center}
\includegraphics[width=4in]{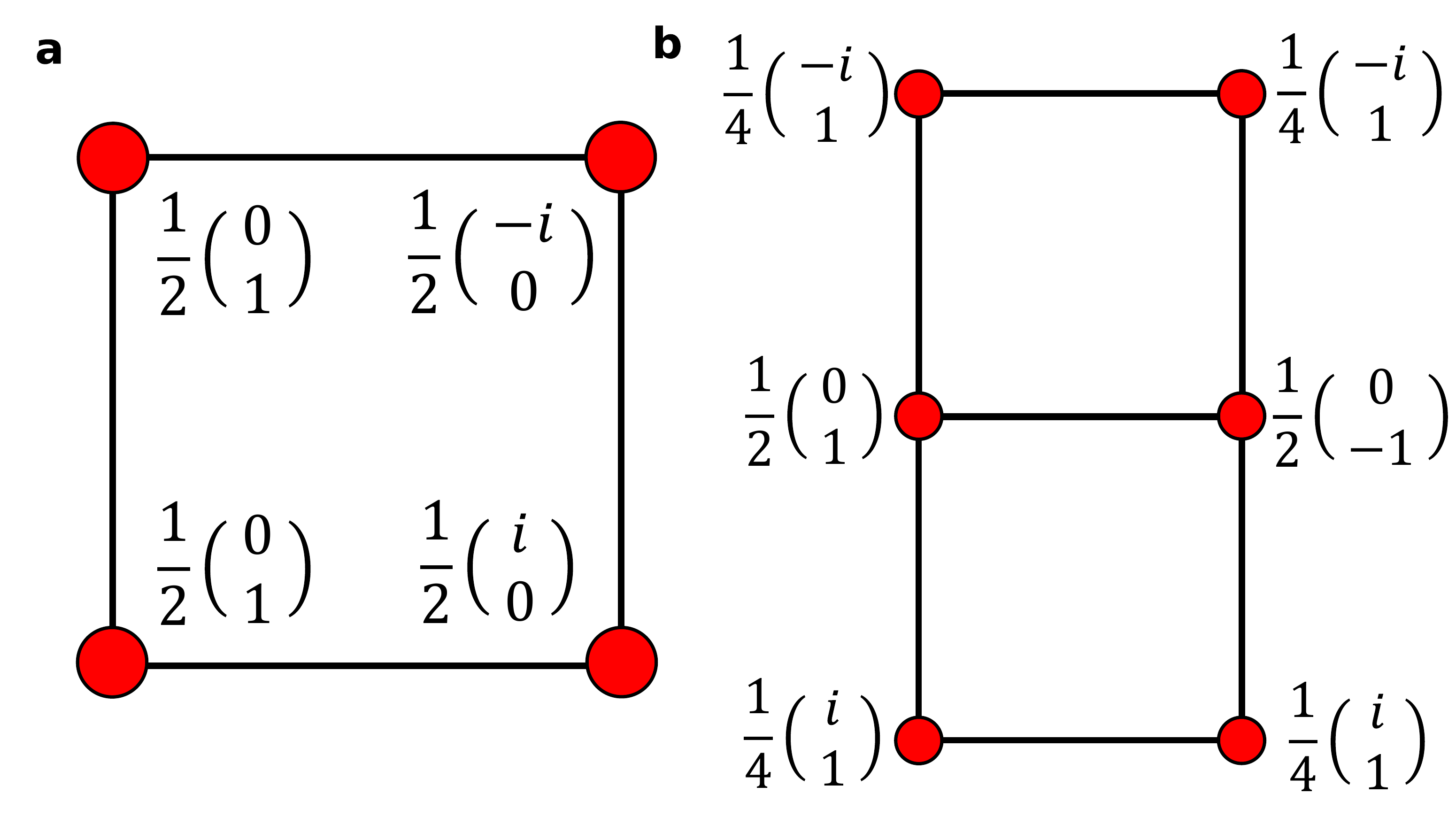}
\caption[]{Supplementary Figure 1. \textbf{Compact valence band Wannier functions for the $n=1$ Chern dartboard insulators.} Each atom is assigned with a $s$ and a $p_y$ orbital, which correspond to the first and the second component of the spinors. $\textbf{\textsf{a}}$ Type I $n=1$ Chern dartboard insulator in the compact limit. The Wannier function transforms as a $p_y$ orbital at Wyckoff position $b$. $\textbf{\textsf{b}}$ Type II $n=1$ Chern dartboard insulator in the compact limit. The Wannier function transforms as a $p_y$ orbital at Wyckoff position $a$.}
\label{compact}
\end{center}
\end{figure}

The polarization and the Berry curvature can be analytically obtained:
\begin{equation}
\begin{split}
P_x&=\int^{2\pi}_0 A_xdk_x=\frac{1-\cos k_y}{2},\\ 
P_y&=\int^{2\pi}_0 A_ydk_y=\frac{1}{2},\\
F_{xy}&=-\frac{\sin k_y}{2},
\end{split}
\end{equation}
which are written in the concise expressions due to the constant normalization of the valence state. The quantized polarization $P_y=1/2$ because of the mirror symmetry $\mathcal{M}_y$ as $P_y\rightarrow-P_y=-1/2=1/2$ modulo integer under mirror reflection. In contrast, the polarization $P_x$ shows the returning Thouless pump (RTP) behavior. The RTP invariant is associated with the $k_x$-dependent $U(1)$ phases at the HSLs arising from the quantized reduced Chern number. Note that it is impossible to find the periodic and smooth gauge that has no $k_x$-dependent $U(1)$ phase at both HSLs. For example, under the gauge of Eq.~(\ref{$n=1$ CDI state}), the valence state at the HSLs is given by
\begin{equation}
\begin{split}
|u^I_v(k_x,k_y=0)\rangle&=\left(\begin{matrix}0\\1\end{matrix}\right),\\
|u^I_v(k_x,k_y=\pi)\rangle&=e^{-ik_x}\left(\begin{matrix}i\\0\end{matrix}\right).
\end{split}
\end{equation}
One can find that at the HSL $k_y=\pi$, $|u^I_v(k_x)\rangle\propto e^{-ik_x}$ which leads to the quantized RTP invariant. Although we can simply remove this gauge by multiplying a phase $e^{ik_x}$ in all the BZ, the valence state at the HSL $k_y=0$ develops a new phase $e^{ik_x}$ such that the difference $\Delta P_x=P_x(k_y=\pi)-P_x(k_y=0)$ is still quantized to one. Therefore, the RTP invariant and the reduced Chern number $\mathcal{C}_1$ are indeed quantized under mirror symmetry.

Notice that here we use the convention that both the Hamiltonian and the intra-cell state are periodic in the BZ. It is interesting to ask what will happen if we use another convention that is more convenient for getting the polarizations and Wannier centers of the systems. In this convention, the valence state has an additional phase $e^{-i\bm{k}\cdot\bm{t}_{\alpha}}$ depending on the basis vector $\bm{t}_{\alpha}$. The intra-cell state and the Hamiltonian are thus aperiodic. To respect the mirror symmetry $\mathcal{M}_y=\sigma_z$ in real space, $\bm{t}_{\alpha}\propto\hat{x}$. Here, we assume the first orbital is localized at $x=0$ and the second orbital is localized at $x=\beta$, where $\beta$ is a constant, and we have set the lattice constant $a=1$. By plugging the phase $e^{-i\beta k_x}$ into the second orbital in Eq.~(\ref{$n=1$ CDI state}), one can easily check that
\begin{equation} \label{nconv}
\begin{split}
P_x&=\frac{(1+\beta)-(1-\beta)\cos k_y}{2},\\ 
P_y&=\frac{1}{2},\\
F_{xy}&=-(1-\beta)\frac{\sin k_y}{2},\\
\mathcal{C}_1&=1-\beta.
\end{split}
\end{equation}
Therefore, the reduced Chern number and the RTP invariant are no longer quantized if we use the aperiodic convention and the two orbitals are not overlapped with each other in the $x$-direction. This actually makes sense: at $k_y=0$, the valence state is spanned by the second orbital at $x=\beta$; and at $k_y=\pi$, the valence state is spanned by the first orbital at $x=1$, see Eq.~(\ref{nconv}). In fact, this is just a demonstration of the violation of the iso-orbital condition that ensures the quantization of the RTP invariant. The iso-orbital condition states that the occupied states have the same symmetry eigenvalues as the band representation induced from one of the basis orbitals, which is obviously violated in the Type I CDI$_1$. Moreover, when $\beta=1$, where all the invariants except $P_y$ become trivial, the valence state in Eq.~(\ref{$n=1$ CDI state}) becomes
\begin{equation}
\begin{split}
|u^I_v(\bm{k},\beta=1)\rangle&=e^{-ik_y/2}\left(\begin{matrix}-e^{-ik_x}\sin\frac{k_y}{2}\\e^{-ik_x}\cos\frac{k_y}{2}\end{matrix}\right) \nonumber\\
&=e^{-ik_y/2}\left(\begin{matrix}-\sin\frac{k_y}{2}\\\cos\frac{k_y}{2}\end{matrix}\right).
\end{split}
\end{equation}
Notice that the $e^{-ik_x}$ phase disappears in the valence state. The corresponding Hamiltonian is given by:
\begin{eqnarray}
H^I_{1}(\bm{k},\beta=1)=\sin k_y\sigma_x+\cos k_y\sigma_z.
\end{eqnarray}
This describes the compact limit of the Su-Schrieffer-Heeger (SSH) model (up to a unitary transformation) stacked in two dimensions. Therefore, the system can be continuously tuned to the 2D weak TI if one allows to move the orbitals in the $x$-direction.

\textbf{Type II CDI\bm{$_1$}.} The compact limit of the Type II CDI$_1$ with $\mathcal{C}_1=1$ is:
\begin{equation}
H^{II}_1(\bm{k})=\frac{1}{2}(1+\cos k_x)\sin 2k_y\sigma_x+\sin k_x\sin k_y\sigma_y+\frac{1}{2}[(1+\cos k_x)(\cos 2k_y-1)+2]\sigma_z.
\end{equation}
One can simply check that the model has two flat bands $E(\bm{k})=\pm|\bm{d}(\bm{k})|=\pm 1$. The normalized valence band state in the compact limit is given by:
\begin{equation}
|u^{II}_v(\bm{k})\rangle=e^{-ik_x/2}\left(\begin{matrix}-\cos\frac{k_x}{2}\sin k_y\\\cos\frac{k_x}{2}\cos k_y+i\sin\frac{k_x}{2}\end{matrix}\right).
\end{equation}
The compact Wannier function is shown in Supplementary Fig.~\ref{compact}b, which occupies six atomic lattice sites. Notice that the Wannier center is localized at Wyckoff position $a$, which is consistent with the symmetry representation. The corresponding polarization and Berry curvature are:
\begin{equation}
\begin{split}
P_x&=\frac{1-\cos k_y}{2}, \\ 
P_y&=0, \\
F_{xy}&=-(1+\cos k_x)\frac{\sin k_y}{2}.
\end{split}
\end{equation}
Here, the bulk polarization $P_y=0$ contrary to the Type I CDI$_1$ which has $P_y=1/2$. The Type II CDI$_1$ shows the same RTP behavior as the Type I CDI$_1$. As a simple check, one can find that the valence state at the HSLs are:
\begin{equation}
\begin{split}
|u^I_v(k_x,k_y=0)\rangle&=\left(\begin{matrix}0\\1\end{matrix}\right),\\
|u^I_v(k_x,k_y=\pi)\rangle&=e^{-ik_x}\left(\begin{matrix}0\\-1\end{matrix}\right).
\end{split}
\end{equation}
Notice the difference between the valence state in the Type I CDI$_1$. Here, they have the same mirror representation at both HSLs. Nevertheless, the appearance of the phase $e^{-ik_x}$ is essential to lead to the quantized reduced Chern number and the RTP invariant. 

If we use the aperiodic convention for the intra-cell states, the second orbital should multiply a phase $e^{-i\beta k_x}$, where $\beta$ is the position of the second orbital in the $x$-direction. Here, we see that the Type II CDI$_1$ is fundamentally different from the Type I CDI$_1$. By multiplying the phase in the second orbital, the valence states at both HSLs also multiply the same phase. Although the polarization $P_x$ get shifted by $\beta$, the RTP invariant stays the same since the difference doesn't change. Therefore, the reduced Chern number also stays the same. This just reflects the fact that the valence state at both HSLs correspond to the same basis orbital $p_y$. Therefore, by shifting its position, the difference between the polarizations at both HSLs cannot change. In fact, since the Type II CDI$_1$ has the same mirror representation at both HSLs, the iso-orbital condition is satisfied.

\textbf{Compact edge theory.} It is also inspiring to discuss the edge physics in the compact limit which has zero correlation length. Since the correlation function is strictly local and becomes zero outside a finite domain, it is possible to find the set of compactly supported eigenstates inside the bulk:
\begin{equation} \label{bulkcompact}
\hat{H}|W^n_{\bm{R}}\rangle=E^n|W^n_{\bm{R}}\rangle,
\end{equation}
where $|W^n_{\bm{R}}\rangle$ are the compactly supported Wannier functions, and the energy $E^n$ is the same for all the Wannier functions within the same band $n$. Since all the eigenstates are degenerate within the same band in the compact limit, the Wannier functions are also the eigenstates of the Hamiltonian. Note that in general Wannier functions are not the eigenstates of the real-space Hamiltonian. Eq.~(\ref{bulkcompact}) is only true in the compact limit. Now we consider the boundaries of the system. Since the correlation function is strictly local, the edge states are also strictly local in the direction normal to the edge:
\begin{eqnarray} \label{edgecompact}
\hat{H}_e(k_e)|e_i(k_e)\rangle=E_e(k_e)|e_i(k_e)\rangle.
\end{eqnarray}
Here, $k_e$ is the reciprocal vector in the direction that parallels to the edge as we assume the periodic boundary condition along that direction. $i$ labels the different edge states that are strictly localized on the boundary, that is, $\langle R'j|e^n_i(k_e)\rangle=0$ for $|R'-R_e|\ge g_e$, where $R'$ is the projected atomic position in the direction normal to the edge, and $R_e$ is the projected position of the edge. $g_e$ is a positive integer.

Now, we can apply the strictly local condition to the Type I CDI$_1$s in the compact limit. First we observe that the edge states that parallel to the $x$-axis have flat energy band due to the quantized bulk polarization $P_y=1/2$. This shares the same characteristics with the 2D weak TIs. However, the edge states that parallel to the $y$-axis are much different due to the quantized reduced Chern number. One can expect that for $k_y\in[0,\pi]$, the edge behaves like a chiral gapless edge states and have the opposite chirality for $k_y\in[\pi,2\pi]$. By observing that the Wannier functions occupy two atomic lattice sites along the $x$-direction (Supplementary Fig.~\ref{compact}), we conclude that the edge states occupy only one atomic lattice site normal to the boundary, that is, $g_e=1$. Therefore, the edge Hamiltonian in Eq.~(\ref{edgecompact}) is exactly given by:
\begin{equation} \label{$n=1$ CDI edgeHamiltonian}
H_e(k_y)=\cos k_y\sigma_z.
\end{equation}
The edge states are exactly given by:
\begin{equation}
\begin{split}
|e^L\rangle&=\left(\begin{matrix}1\\0\end{matrix}\right),\\
|e^R\rangle&=\left(\begin{matrix}0\\1\end{matrix}\right),
\end{split}
\end{equation}
with energy $E^L_e(k_y)=\cos k_y$ and $E^R_e(k_y)=-\cos k_y$. Here the superscript $L$ and $R$ represent the edge at the left or right boundary. Although the edge Hamiltonian is a two-band Hamiltonian, the two edge states are spatially living in the opposite boundaries. Since the edge states need to be orthogonal to the bulk states, there is only one edge state living in each boundary. One can observe that indeed the edge states flow between the bulk bands with energy $E(\bm{k})=\pm 1$ for $k\in[0,\pi]$ and opposite for $k\in[\pi,2\pi]$. Recall from the argument in the main text, this connectivity persists as long as mirror symmetry is preserved and the sharp boundary condition is assumed, even though the system is not in the compact limit. In addition, the Hamiltonian Eq.~(\ref{$n=1$ CDI edgeHamiltonian}) describes the simplest one-dimensional conducting chain with nearest-neighbor coupling. 

The appearance of the edge states can be explained through the angular-momentum anomaly:
\begin{eqnarray} \label{anomaly}
&&-\sharp_f\mathcal{L}_v(k_y=0)+\sharp_f\mathcal{L}_v(k_y=\pi) \nonumber\\
&&=\sharp_f\mathcal{L}_c(k_y=0)-\sharp_f\mathcal{L}_c(k_y=\pi)=\Delta P_x.
\end{eqnarray}
Here $\sharp_f\mathcal{L}_i$ denotes the number of left-edge-localized states with the same mirror representations of the valence bands $i=v$ or conduction bands $i=c$. It is easy to see that $\sharp_f\mathcal{L}_v(k_y=0)=0$ and $\sharp_f\mathcal{L}_v(k_y=\pi)=1$, in addition with $\sharp_f\mathcal{L}_c(k_y=0)=1$ and $\sharp_f\mathcal{L}_c(k_y=\pi)=0$. Since the RTP invariant is a topological invariant protected by mirror symmetry, there must exist edge-localized states such that the relation is obeyed. The edge spectrum flows between the bulk bands since they have different mirror representations at $k_y=0,\pi$ with the valence or conduction bands. Moreover, the edge states at the left and right boundaries form a conjugate pair as their mirror representations are opposite, and the angular-momentum anomaly is satisfied with an opposite sign.

We again investigate the edge physics for the Type II CDI$_1$. First, consider the edge states that parallel to the $y$-axis. In this case, one can expect the similar edge behavior as the Type I CDI$_1$s. Since the compact Wannier functions also occupy two atomic lattice sites along the $x$-direction (Supplementary Fig.~\ref{compact}), we have $g_e=1$, and the edge Hamiltonian is exactly given by:
\begin{equation}
H_e(k_y)=\frac{1}{2}\sin 2k_y\sigma_x+\frac{1}{2}(\cos 2k_y+1)\sigma_z.
\end{equation}
Here, contrary to the previous case, the edge Hamiltonian is not a simple one-dimensional conducting with nearest-neighbor coupling but rather the phase transition point of the SSH model between winding number $w=0$ and $w=2$. This is consistent with the intuition as the Type II CDI$_1$ has more robust topology than the Type I CDI$_1$. The normalized edge states are given by:
\begin{equation}
\begin{split}
|e^L\rangle&=\frac{1}{2}\left(\begin{matrix}1+e^{-ik_y}\\-i+ie^{-ik_y}\end{matrix}\right) \\
|e^R\rangle&=\frac{1}{2}\left(\begin{matrix}i-ie^{-ik_y}\\1+e^{-ik_y}\end{matrix}\right),
\end{split}
\end{equation}
with energy $E^L_e(k_y)=\cos k_y$ and $E^R_e(k_y)=-\cos k_y$. The edge states flow between the bulk bands with energy $E(\bm{k})=\pm 1$ for $k\in[0,\pi]$ and opposite for $k\in[\pi,2\pi]$. Therefore, the bulk-boundary correspondence $N^+_e=N^-_e=\mathcal{C}_1$ is satisfied. We emphasize again that the spectral flow is protected even though the system is not in the compact limit. Here, the Type II CDI$_1$s host nontrivial one-dimensional conducting chain at the edges. Since the bulk bands have the same mirror eigenvalues at $k_y=0,\pi$, the edge states must have different mirror eigenvalues at $k_y=0,\pi$, contrary to the Type I CDI$_1$s where the edge states have the same mirror eigenvalues. We can also check that the angular-momentum anomaly is obeyed from Eq.~(\ref{anomaly}), and the edge states at the left and right boundaries also form a conjugate pair.

The edge states that parallel to the $x$-axis in the compact limit show quite different edge physics. Here, we expect $g_e=2$ since the Wannier functions occupy three atomic lattice sites along the $y$-direction (Supplementary Fig.~\ref{compact}). The edge Hamiltonian can thus be exactly obtained:
\begin{equation}
H_e(k_x)=-\frac{1}{2}\sin k_x\tau_y\sigma_y+\frac{1}{2}(1-\cos k_x)\tau_0\sigma_z.
\end{equation}
Here, $\tau_i$ represents the Pauli matrices acting on the sublattices since now we have two atomic lattice sites $g_e=2$ for the edge states. $\tau_0=I$ is the identity acting on the sublattices. The mirror symmetry representation becomes $\mathcal{M}_{ye}=\tau_x\sigma_z$, and it acts as an internal symmetry for the edge states. The edge Hamiltonian describes the phase transition point of the four-band SSH model. By explicitly solving the normalized edge states, we obtain:
\begin{equation}
\begin{split}
|e^U_1\rangle&=\frac{1}{2}\left(-i\sin\frac{k_x}{2}-i,\sin\frac{k_x}{2}-1,\cos\frac{k_x}{2},-i\cos\frac{k_x}{2}\right)^T \\
|e^D_1\rangle&=\mathcal{M}_{ye}|e^U_1\rangle \\
|e^U_2\rangle&=\frac{1}{2}\left(i\sin\frac{k_x}{2}-i,-\sin\frac{k_x}{2}-1,-\cos\frac{k_x}{2},i\cos\frac{k_x}{2}\right)^T \\
|e^D_2\rangle&=\mathcal{M}_{ye}|e^U_2\rangle,
\end{split}
\end{equation}
with energy $E^U_1=E^D_1=\sin (k_x/2)$, $E^U_2=E^D_2=-\sin (k_x/2)$, where the superscripts label the edge states at the upper or lower boundary, which are related by mirror symmetry.

One may immediately notice that the edge states form the Möbius fermions as discussed in the main text. Apparently, both the edge states $|e^U_1\rangle$ and $|e^U_2\rangle$ are not periodic in the edge BZ even though we use the periodic convention. However, notice that  $|e^U_1(k_x=2\pi)\rangle=|e^U_2(k_x=0)\rangle$, $|e^U_1(k_x=0)\rangle=|e^U_2(k_x=2\pi)\rangle$ and the two edge states together form a periodic and smooth function between $k_x\in[0,4\pi]$. The corresponding energies $E^U_1(k_x)$ and $E^U_2(k_x)$ also show the Möbius twist.

\addcontentsline{toc}{section}{Supplementary Note 3. Type II $n=3$ Chern dartboard insulator}
\section*{\centering Supplementary Note 3. Type II $\bm{n=3}$ Chern dartboard insulator} \label{sec3}

\begin{figure}
\begin{center}
\includegraphics[width=4in]{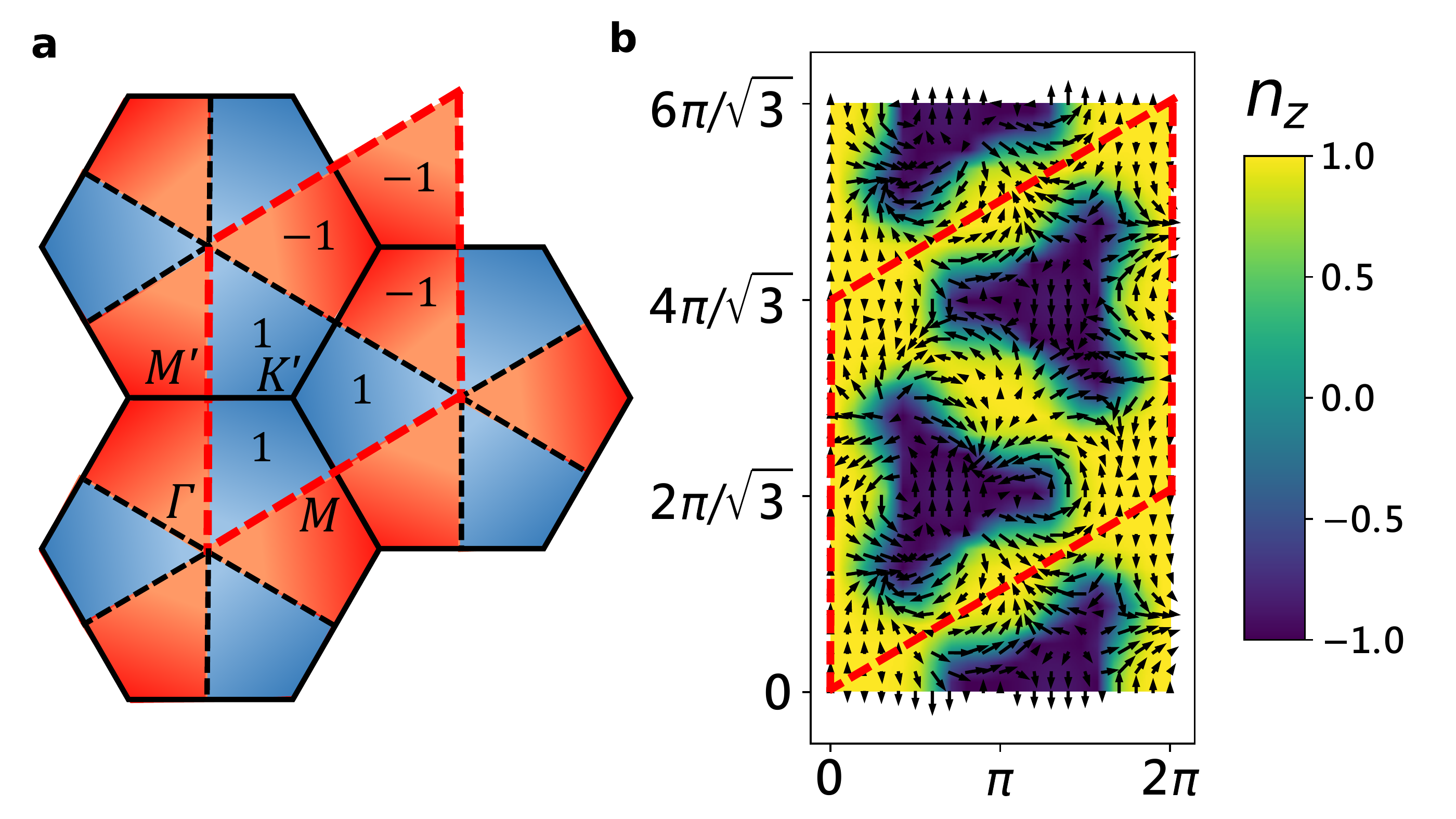}
\caption[]{Supplementary Figure 2. \textbf{Type II $\bm{n=3}$ Chern dartboard insulator.} $\textbf{\textsf{a}}$ The quantized Chern number inside the irreducible BZs. The dashed lines indicate the HSLs. The reciprocal lattice vectors $\bm{G}_1=(2\pi,2\pi/\sqrt{3})^T$ and $\bm{G}_2=(0,4\pi/\sqrt{3})^T$ form the rhombus-shape BZ, which is indicated by the red dashed lines. The numbers denote the Chern numbers inside the irreducible BZs, enclose by the path $\Gamma-M'-K'-M-\Gamma$. $\textbf{\textsf{b}}$ The peudospin texture of the Type II CDI$_3$. There are totally $6$ skyrmions and $3$ antiskyrmions inside half of the BZ, which add up to $\mathcal{C}_1=6-3=3$. The antiskyrmions are not clear in this figure since they are very close to the skyrmions. Here $m=1.5$.}
\label{n=3-2}
\end{center}
\end{figure}

The Type II CDI$_3$ has a quantized reduced Chern number $\mathcal{C}_3$ inside $1/6$ of the BZ, see Supplementary Fig.~\ref{n=3-2}a. It can be built by a $s$ and a $f_{x(x^2-3y^2)}$ orbital on the triangular lattice with lattice vectors $\bm{a}_1=(1,0)^T$, $\bm{a}_2=(1/2,\sqrt{3}/2)^T$. The $f_{x(x^2-3y^2)}$ is odd under three mirror symmetries: $\mathcal{M}_x$, $C_3\mathcal{M}_xC_3^{-1}$, $C_3^2\mathcal{M}_xC_3^{-2}$. All these mirror symmetries have the same representation $\sigma_z$. The $C_3$ rotation has a trivial representation $C_3=I$ since it can be obtained by a combination of two mirror symmetries. The model with $\mathcal{C}_3=1$ is:
\begin{equation}
\begin{split}
H^{II}_3(\bm{k})=\big[&\sin \frac{5}{2}k_x\sin \frac{\sqrt{3}}{2}k_y-\sin 2k_x\sin \sqrt{3}k_y+\sin \frac{1}{2}k_x\sin \frac{3\sqrt{3}}{2}k_y\big]\sigma_x \\
+\big[&\sin \frac{5}{2}k_x\cos \frac{\sqrt{3}}{2}k_y-\sin 2k_x\cos \sqrt{3}k_y-\sin \frac{1}{2}k_x\cos \frac{3\sqrt{3}}{2}k_y\big]\sigma_y \\
+\bigg[&m+\sum_{a=1}^6\big(e^{i\bm{t}_1(a)\cdot\bm{k}}+e^{2i\bm{t}_1(a)}\big)\bigg]\sigma_z,
\end{split}
\end{equation}
where $\bm{t}_1(a)=[\cos(\pi a/3),\sin(\pi a/3)]^T$. The $\sigma_z$ term contains the hoppings with range $\bm{a}_1$, $2\bm{a}_1$, and also the ones generated by all the $C_6$ rotations. The $\sigma_x$ and $\sigma_y$ contains the hoppings with range $2\bm{a}_1+\bm{a}_2$ and the ones generated by three mirror symmetries. The system has a quantized reduced Chern number $\mathcal{C}_3=1$ if $0.5<m<2.5$. The pseudospin texture is plotted in Supplementary Fig.~\ref{n=3-2}b.

\begin{figure}
\begin{center}
\includegraphics[width=5in]{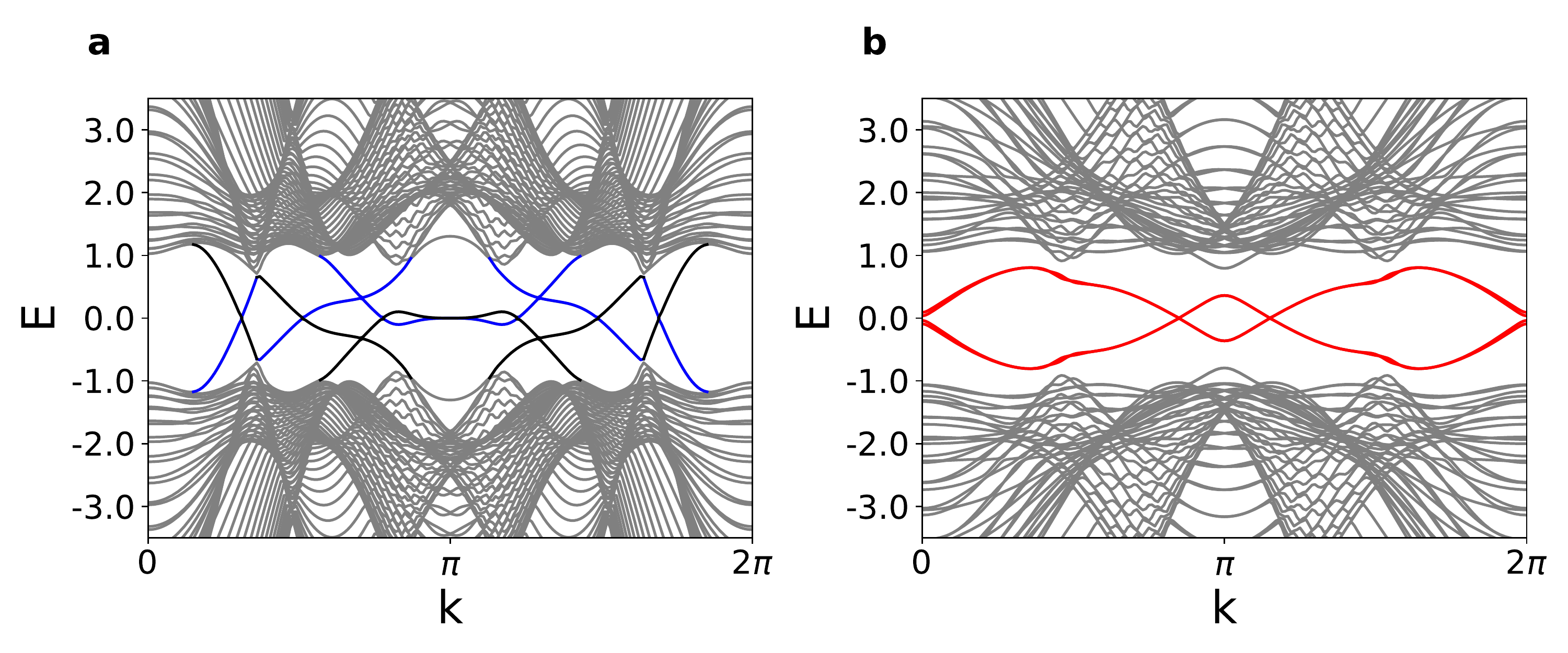}
\caption[]{Supplementary Figure 3. \textbf{Nanoribbon band structures of the Type II $\bm{n=3}$ Chern dartboard insulator.} $\textbf{\textsf{a}}$ The spectrum in the $x$-direction. The edge states localized at the lower (upper) edge are denoted by the blue (black) color. $\textbf{\textsf{b}}$ The spectrum in the $y$-direction. In this case, all the edge states are doubly degenerate because of the mirror symmetry $\mathcal{M}_x$, as highlighted by the red color. Here $m=1.5$.}
\label{n=3-2edge}
\end{center}
\end{figure}

The Type II CDI$_3$ is actually a little different from other CDIs. The sBZs with quantized reduced Chern number are not entirely enclosed by the HSLs. Nevertheless, the sBZ topology is still well-defined. This comes from the fact that the valence states inside the half rhombus-shape BZ, which is an equilateral triangle, is $C_3$-symmetric around the center $K$ or $K'$. This emergent $C'_3$ symmetry is a combination of a $C_3$ symmetry plus a $\bm{k}$-space translation. Next, we use the Stokes theorem:
\begin{equation}
\frac{1}{2\pi}\oint\bm{A}\cdot d\bm{k}=\frac{1}{2\pi}\int_\text{sBZ} d^2k F_{xy},
\end{equation}
where the loop integral is around the path $\Gamma-M'-K'-M-\Gamma$, see Supplementary Fig.~\ref{n=3-2}a. Due to the $C'_3$ symmetry, the path integral:
\begin{equation}
\frac{1}{2\pi}\int_{M'}^{K'}\bm{A}\cdot d\bm{k}+\frac{1}{2\pi}\int_{K'}^{M}\bm{A}\cdot d\bm{k}=0,
\end{equation}
and the remained integral:
\begin{equation}
\frac{1}{2\pi}\int_{\Gamma}^{M'}\bm{A}\cdot d\bm{k}+\frac{1}{2\pi}\int_{M}^{\Gamma}\bm{A}\cdot d\bm{k}=-\mathcal{C}'_3\in\mathbb{Z}
\end{equation}
is a loop integral since $M'$ maps to $M$ by $C'_3$ symmetry, which has a trivial representation. Therefore, the region enclosed by this path has a quantized reduced Chern number $\mathcal{C}'_3\in\mathbb{Z}$. 

In Supplementary Fig.~\ref{n=3-2edge}, we plot the edge states for the Type II CDI$_3$. Notice that the edge states in the $y$-direction are not connecting with the bulk bands and thus they show the feature of Möbius fermions.  For both the zigzag and flat edges, we observe that the bulk-boundary correspondence is still satisfied for each edge:
\begin{eqnarray}
N^+_e=N^-_e=3=n\mathcal{C}_n.
\end{eqnarray}

\addcontentsline{toc}{section}{Supplementary Note 4. Noncompact Atomic Insulators}
\section*{\centering Supplementary Note 4. Noncompact Atomic Insulators} \label{sec4}

Contrary to the CDI$_1$s, the $n=2,4,6$ CDIs are noncompact atomic insulators. This comes from the fact that the non-$\delta$ compact Wannier functions are incompatible with $C_2$ symmetry of trivial representation $C_2=I$. If the Wannier functions are compact, there exists a minimal circle with radius $r_m$ such that $\langle \bm{R}'j|W_{\bm{r}_0}\rangle=0$ for $|\bm{R}'-\bm{r}_0|>r_m$. Here, $|\bm{R}'j\rangle$ represents the atomic lattice state at position $\bm{R}'$ and $j\in 1,2,..,N$ denotes the orbitals. $\bm{r}_0$ denotes the center of the circle and the Wannier function $|W_{\bm{r}_0}\rangle$. Let us define the farthest basis state collection $\{|\bm{R}_Mj\rangle\}$ with nonzero overlap $\langle \bm{R}_Mj|W_{\bm{r}_0}\rangle\neq 0$ for at least one $j$,  that has the maximal distance from the center of the circle $|\bm{R}_M-\bm{r}_0|=r_m\ge|\bm{R}'-\bm{r}_0|$ for all $\langle \bm{R}'j|W_{\bm{r}_0}\rangle\neq 0$ for at least one $j$. 

\begin{figure}
\begin{center}
\includegraphics[width=4in]{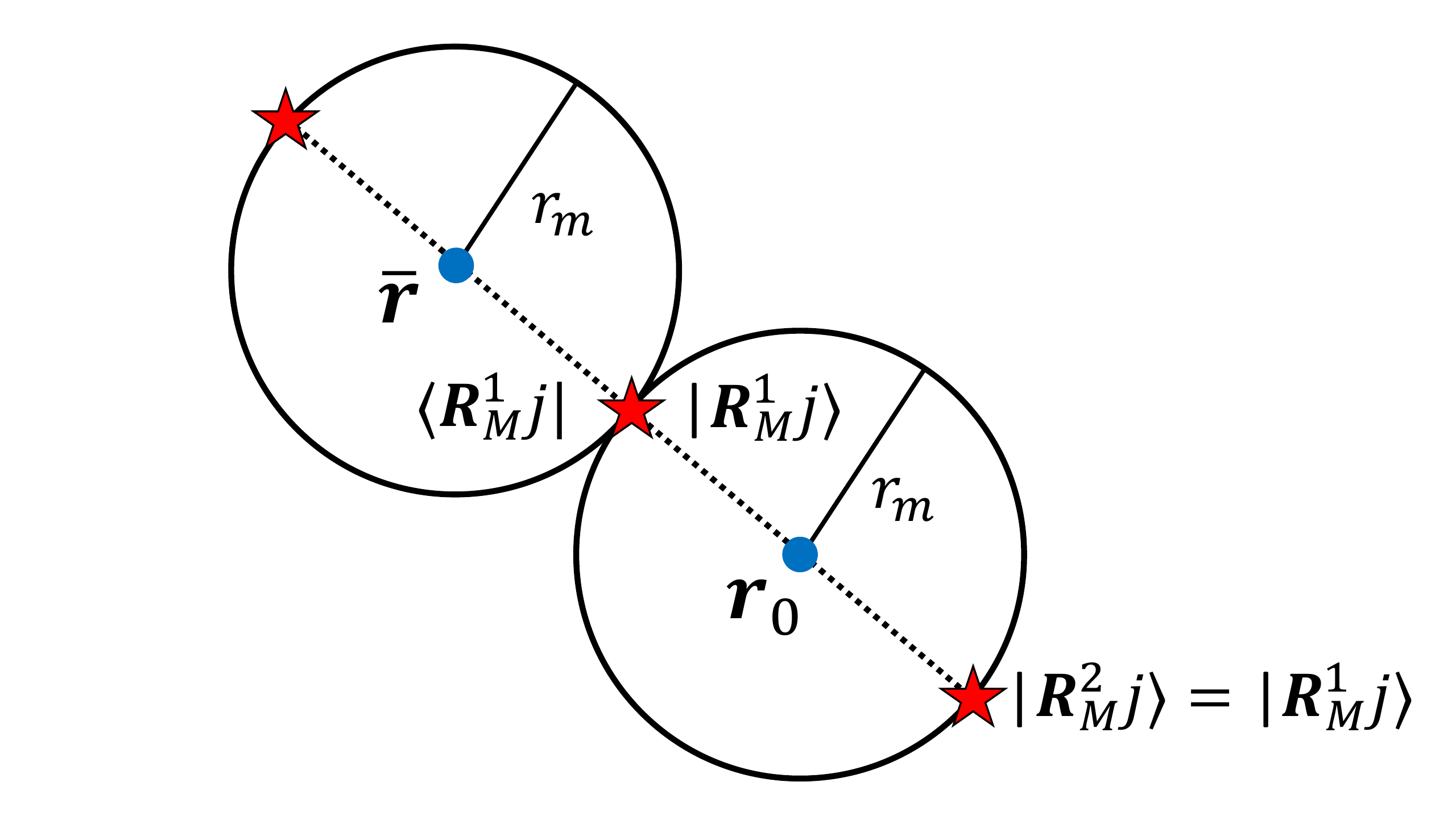}
\caption[]{Supplementary Figure 4. \textbf{Proof for the noncompact atomic insulators.} The key point is that the Wannier function centered at $\bar{\bm{r}}=2\bm{R}^1_M-\bm{r}_0$ must have a nonzero overlap $|\langle\bm{R}^1_Mj|W_{\bm{r}_0}\rangle|^2$ with the Wannier function centered at $\bm{r}_0$.}
\label{noncompact}
\end{center}
\end{figure}

Now we arbitrarily choose a farthest basis state $|\bm{R}^1_Mj\rangle$ from the collection $\{|\bm{R}_Mj\rangle\}$. If this basis state doesn't lie on the center of the Wannier function $\bm{r}_0$, there exists another farthest basis state $|\bm{R}^2_Mj\rangle=|(2\bm{r}_0-\bm{R}^1_M)j\rangle=C_2|\bm{R}^1_Mj\rangle=|\bm{R}^1_Mj\rangle$ related by a $C_2$ rotation, see Fig.~\ref{noncompact}. However, if we consider another Wannier function centered at $\bar{\bm{r}}=2\bm{R}^1_M-\bm{r}_0$, the overlap between the two Wannier functions is given by:
\begin{equation}
\begin{split}
\langle W_{\bar{\bm{r}}}|W_{\bm{r}_0}\rangle&=\sum_{\bm{R}',j}\langle W_{\bar{\bm{r}}}|\bm{R}'j\rangle\langle\bm{R}'j|W_{\bm{r}_0}\rangle \nonumber\\
&=|\langle\bm{R}^1_Mj|W_{\bm{r}_0}\rangle|^2\neq0,
\end{split}
\end{equation}
for at least one $j$. Notice that here we have used the equal condition of the triangular inequality:
\begin{equation} \label{inequality}
|\bm{R}'-\bm{r}_0|+|\bm{R}'-\bar{\bm{r}}|\ge |\bar{\bm{r}}-\bm{r}_0|=2r_m.
\end{equation}
Since $\langle \bm{R}'j|W_{\bm{r}_0}\rangle=0$ for $|\bm{R}'-\bm{r}_0|>r_m$ and $\langle \bm{R}'j|W_{\bar{\bm{r}}}\rangle=0$ for $|\bm{R}'-\bar{\bm{r}}|>r_m$, the only overlap between the two Wannier functions is at $\bm{R}'=\bm{R}^1_M$. However, this overlap must be nonzero due to the trivial representation of $C_2$ symmetry, and the Wannier functions are not orthogonal to each other. Therefore, the non-$\delta$ orthonormal compact Wannier functions are incompatible with a $C_2$ rotation of trivial representation. The only loophole of this proof is that the minimal circle has radius $r_m=0$ such that there is only one farthest basis state, which is just the Wannier function itself. Therefore, it is possible to have the $\delta$-like compact Wannier function, as it should be.

For the CDIs protected by three mirror symmetries, we cannot use the above proof since there is no $C_2$ symmetry in the systems. Therefore, we explicitly search for different shapes of symmetric compact Wannier functions that satisfy the orthogonality. We find that the compact Wannier functions less or equal than 25 atomic sites are all inconsistent with the symmetries and orthogonality conditions. For simplicity, we consider the two-band models built on the triangular lattice with primitive lattice vectors $\bm{a}_1=(1,0)$ and $\bm{a}_2=(1/2,\sqrt{3}/2)$ in the unit of a lattice constant. Then orthogonality conditions can be written as:
\begin{equation}
\langle W_{\bm{R}}|W_{\bm{R'}}\rangle=\delta_{\bm{RR'}},
\end{equation}
for any $\bm{R}$ and $\bm{R'}$. Mirror symmetric conditions can be written as:
\begin{equation}
\mathcal{M}_i|W_{\bm{R}}\rangle=\pm|W_{\bm{R}}\rangle.
\end{equation}
Here, the real-space mirror symmetry operator $\mathcal{M}_i=R_i\otimes\sigma_z$, where $R_i$ represents mirror reflection in real space with $i=1,2,3$, and $\sigma_z$ is the mirror symmetry representation acting on the orbital basis. Notice that the systems also have $C_3=I$ symmetry arising from three mirror symmetries. Without loss of generality, we consider only the cases that the compact Wannier function transforms as a $s$ orbital under mirror symmetries, $\mathcal{M}_i|W_{\bm{R}}\rangle=|W_{\bm{R}}\rangle$. From these symmetry constraints, we can easily list different shapes of the symmetric compact Wannier functions. Then we use all the orthogonality conditions to constrain each component of the compact Wannier functions. We label each configuration with the total number of the occupied atomic sites. The atomic lattice states are denoted as
\begin{equation}
\begin{split}
|i\rangle=\left(\begin{matrix}a_i\\b_i\end{matrix}\right),
\end{split}
\end{equation}
where $i$ labels the atomic lattice sites as shown in Supplementary Fig.~\ref{n=3 compact} and~\ref{n=3-2 compact}. Notice that the $C_3=I$ symmetry substantially reduces the number of different atomic lattice states for each compact Wannier function. In the following, we discuss the Type I and Type II CDI$_3$s respectively.

\textbf{Type I CDI\bm{$_3$}.} The mirror reflection planes pass through the atoms of the triangular lattice, see Supplementary Fig.~\ref{n=3 compact}a. The three mirror symmetries can be denoted as $\mathcal{M}_y,C_3\mathcal{M}_yC_3^{-1},C_3^2\mathcal{M}_yC_3^{-2}$. The basis orbitals consist of a $s$ orbital and a $f_{y(3x^2-y^2)}$ orbital.

\begin{figure}
\begin{center}
\includegraphics[width=5in]{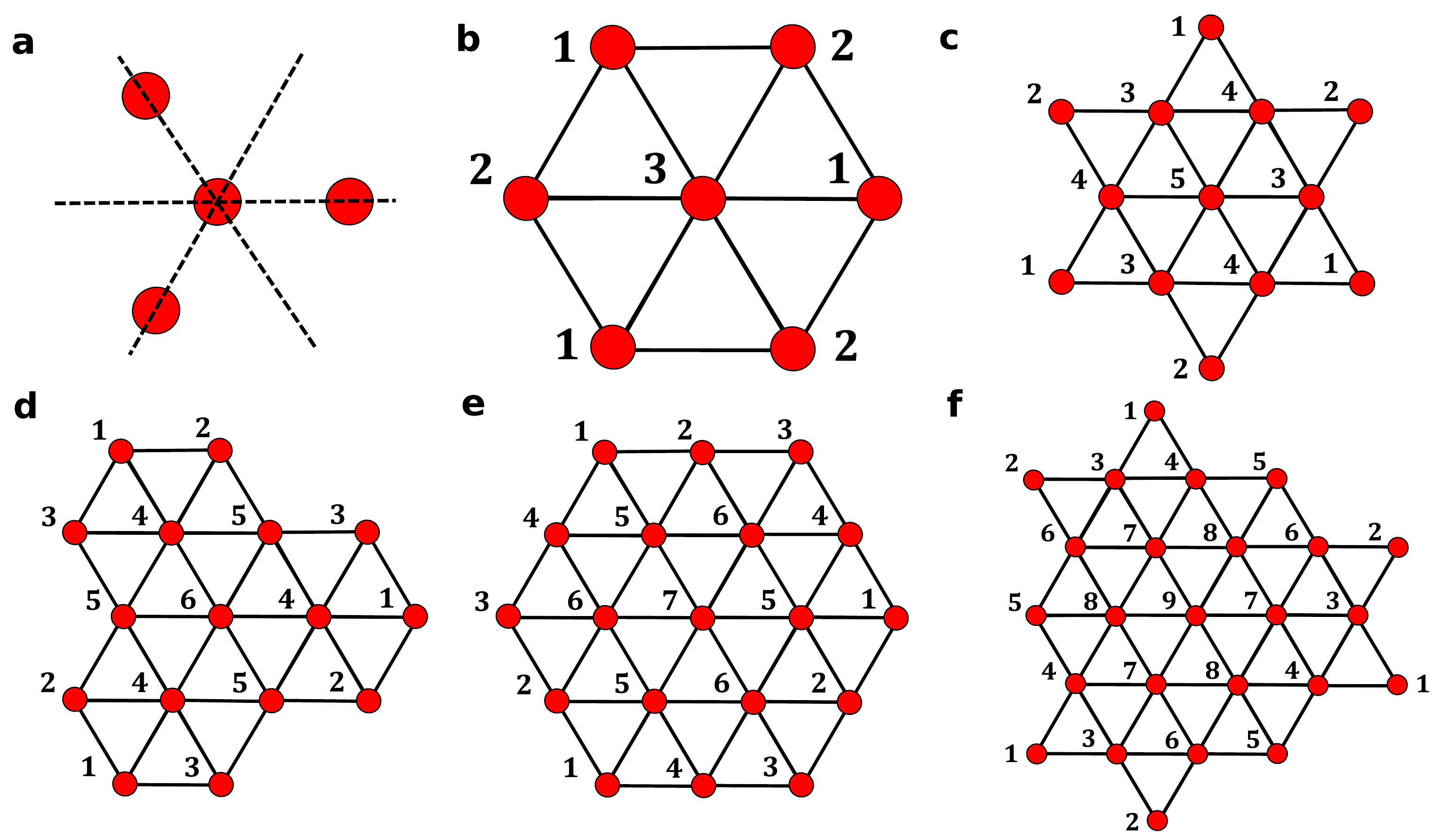}
\caption[]{Supplementary Figure 5. \textbf{Compact Wannier functions for the Type I $\bm{n=3}$ Chern dartboard insulator.} $\textbf{\textsf{a}}$ An equilateral triangular shape of the Wannier function. The dashed lines show the mirror reflection planes that pass through the atomic sites. $\textbf{\textsf{b}}$ $N=7$ compact Wannier function. $\textbf{\textsf{c}}$ $N=13$ compact Wannier function. $\textbf{\textsf{d}}$ $N=16$ compact Wannier function. $\textbf{\textsf{e}}$ $N=19$ compact Wannier function. $\textbf{\textsf{f}}$ $N=25$ compact Wannier function.}
\label{n=3 compact}
\end{center}
\end{figure}

$\bm{N=4}$ \textbf{compact Wannier function}. The configuration of the compact Wannier function is shown in Supplementary Fig.~\ref{n=3 compact}a. Since the mirror reflection planes pass through each atom, the atomic lattice states should be the eigenstates of the mirror symmetry representation $\sigma_z$ with eigenvalue $1$. That is,
\begin{equation}
\begin{split}
\sigma_z|1\rangle=|1\rangle=\left(\begin{matrix}a_1\\0\end{matrix}\right),
\end{split}
\end{equation}
where $|1\rangle$ are the three outer atomic lattice states that constitute a equilateral triangle. The orthogonality condition requires that  
\begin{equation}
\begin{split}
\langle 1|1\rangle=0,
\end{split}
\end{equation}
which is impossible. We note that this contradiction arises as long as the shape of the compact Wannier function is an equilateral triangle regardless of the total number of the occupied atomic lattice sites inside. Therefore, in the following, we have automatically excluded the compact Wannier functions with this kind of shape.

$\bm{N=7}$ \textbf{compact Wannier function}. The configuration of the compact Wannier function is shown in Supplementary Fig.~\ref{n=3 compact}b. First we list the mirror symmetry constraints
\begin{subequations}
\begin{align}
&\sigma_z|1\rangle=|1\rangle=\left(\begin{matrix}a_1\\0\end{matrix}\right), \label{N=7-1}\\
&\sigma_z|2\rangle=|2\rangle=\left(\begin{matrix}a_2\\0\end{matrix}\right), \label{N=7-2}\\
&\sigma_z|3\rangle=|3\rangle=\left(\begin{matrix}a_3\\0\end{matrix}\right). \label{N=7-3}
\end{align}
\end{subequations}
The orthogonality conditions require that,
\begin{subequations}
\begin{align}
&\langle 2|1\rangle=0\Rightarrow a_2^*a_1=0, \label{N=7-4}\\
&\langle 1|1\rangle+\langle 2|2\rangle=0\Rightarrow |a_1|^2+|a_2|^2=0, \label{N=7-5}\\
&\langle 3|1\rangle+\langle 1|2\rangle + \langle 1|2\rangle + \langle 2|3\rangle=0 \label{N=7-6}.
\end{align}
\end{subequations}
From Eq.~(\ref{N=7-5}) we can conclude that $a_1=a_2=0$. Therefore the $N=7$ compact Wannier function is inconsistent.

$\bm{N=13}$ \textbf{compact Wannier function}. The configuration of the compact Wannier function is shown in Supplementary Fig.~\ref{n=3 compact}c. The mirror symmetry constraints are:
\begin{subequations}
\begin{align}
&|2\rangle=\sigma_z|1\rangle\Rightarrow \left(\begin{matrix}a_2\\b_2\end{matrix}\right)=\left(\begin{matrix}a_1\\-b_1\end{matrix}\right), \label{n=13-1}\\
&\sigma_z|3\rangle=|3\rangle=\left(\begin{matrix}a_3\\0\end{matrix}\right), \label{n=13-2}\\
&\sigma_z|4\rangle=|4\rangle=\left(\begin{matrix}a_4\\0\end{matrix}\right), \label{n=13-3}\\
&\sigma_z|5\rangle=|5\rangle=\left(\begin{matrix}a_5\\0\end{matrix}\right). \label{n=13-4}
\end{align}
\end{subequations}
The orthogonality conditions require that,
\begin{subequations}
\begin{align}
&\langle 2|1\rangle=0\Rightarrow |a_1|^2-|b_1|^2=0, \label{n=13-5}\\
&\langle 1|1\rangle+\langle 2|2\rangle=0\Rightarrow |a_1|^2+|b_1|^2+|a_1|^2+|b_1|^2=0, \label{n=13-6}\\
&\langle 4|1\rangle+\langle 2|3\rangle=0, \label{n=13-7}\\
&\langle 3|1\rangle+\langle 3|2\rangle+\langle 4|3\rangle+\langle 1|4\rangle+\langle 2|4\rangle=0, \label{n=13-8}\\
&\langle 5|1\rangle+\langle 1|2\rangle+\langle 3|3\rangle+\langle 4|4\rangle+\langle 1|2\rangle+\langle 2|5\rangle=0, \label{n=13-9}\\
&\langle 4|1\rangle+\langle 4|2\rangle+\langle 5|3\rangle+\langle 3|4\rangle+\langle 3|4\rangle+\langle 4|5\rangle+\langle 1|3\rangle+\langle 2|3\rangle=0. \label{n=13-10}
\end{align}
\end{subequations}
From Eq.~(\ref{n=13-6}), we obtain $a_1=b_1=a_2=b_2=0$ and the Wannier function becomes the $N=7$ compact Wannier function. Therefore the $N=13$ compact Wannier function is inconsistent. It is clear from this case that all the Wannier function with star shape is also inconsistent.

$\bm{N=16}$ \textbf{compact Wannier function}. The configuration of the compact Wannier function is shown in Supplementary Fig.~\ref{n=3 compact}d. The mirror symmetry constraints are:
\begin{subequations}
\begin{align}
&|3\rangle=\sigma_z|2\rangle\Rightarrow \left(\begin{matrix}a_3\\b_3\end{matrix}\right)=\left(\begin{matrix}a_2\\-b_2\end{matrix}\right), \label{n=16-1}\\
&\sigma_z|1\rangle=|1\rangle=\left(\begin{matrix}a_1\\0\end{matrix}\right), \label{n=16-2}\\
&\sigma_z|4\rangle=|4\rangle=\left(\begin{matrix}a_4\\0\end{matrix}\right), \label{n=16-3}\\
&\sigma_z|5\rangle=|5\rangle=\left(\begin{matrix}a_5\\0\end{matrix}\right), \label{n=16-4}\\
&\sigma_z|6\rangle=|6\rangle=\left(\begin{matrix}a_6\\0\end{matrix}\right). \label{n=16-5}
\end{align}
\end{subequations}
The orthogonality conditions require that,
\begin{subequations}
\begin{align}
&\langle 3|1\rangle=0\Rightarrow a_2^*a_1=0,\label{n=16-6}\\
&\langle 1|1\rangle+\langle 3|2\rangle=0\Rightarrow |a_1|^2+|a_2|^2-|b_2|^2=0,\label{n=16-7}\\
&\langle 5|1\rangle+\langle 2|2\rangle+\langle 3|3\rangle=0\Rightarrow a_5^*a_1+|a_2|^2+|b_2|^2+|a_2|^2+|b_2|^2=0,\label{n=16-8}\\
&\langle 4|1\rangle+\langle 5|2\rangle+\langle 1|3\rangle+\langle 3|4\rangle=0\Rightarrow a_4^*a_1+a_5^*a_2+a_1^*a_2+a_2^*a_4=0,\label{n=16-9}\\
&\langle 6|1\rangle+\langle 4|2\rangle+\langle 4|3\rangle+\langle 5|4\rangle+\langle 2|5\rangle+\langle 3|5\rangle=0\Rightarrow a_6^*a_1+a_4^*a_2+a_4^*a_2+a_5^*a_4+a_2^*a_5+a_2^*a_5=0,\label{n=16-10}\\
&\langle 5|1\rangle+\langle 6|2\rangle+\langle 2|3\rangle+\langle 4|4\rangle+\langle 5|5\rangle+\langle 2|3\rangle+\langle 1|5\rangle+\langle 3|6\rangle=0 \nonumber\\
&\Rightarrow a_5^*a_1+a_6^*a_2+|a_2|^2-|b_2|^2+|a_4|^2+|a_5|^2+|a_2|^2-|b_2|^2+a_1^*a_5+a_2^*a_6=0,\label{n=16-11}\\
&\langle 4|1\rangle+\langle 5|2\rangle+\langle 5|3\rangle+\langle 6|4\rangle+\langle 4|5\rangle+\langle 1|3\rangle+\langle 4|5\rangle+\langle 5|6\rangle+\langle 2|4\rangle+\langle 1|2\rangle+\langle 3|4\rangle=0.\label{n=16-12}
\end{align}
\end{subequations}
From Eq.~(\ref{n=16-6}), it is clear that either $a_1=0$ or $a_2=0$. However, $a_1\neq0$ otherwise the Wannier function becomes the $N=13$ compact Wannier function, which is inconsistent. Therefore the only possibility is that 
\begin{equation} \label{n=16-13}
a_2=a_3=0,
\end{equation}
and Eq.~(\ref{n=16-7}) becomes
\begin{equation} \label{n=16-14}
|b_2|^2=|a_1|^2.
\end{equation}
From Eq.~(\ref{n=16-8}), (\ref{n=16-13}) and (\ref{n=16-14}), it follows that
\begin{equation} \label{n=16-15}
a_5^*a_1+2|a_1|^2=0.
\end{equation}
Since $a_1\neq 0$, it implies that
\begin{equation} \label{n=16-16}
a_5=-2a_1.
\end{equation}
From Eq.~(\ref{n=16-9}) and (\ref{n=16-13}), we obtain
\begin{equation} \label{n=16-17}
a_4=0.
\end{equation}
From Eq.~(\ref{n=16-10}), (\ref{n=16-13}) and (\ref{n=16-17}),
\begin{equation} \label{n=16-18}
a_6=0.
\end{equation}
The equation (\ref{n=16-11}) by plugging into Eq.~(\ref{n=16-13}), (\ref{n=16-14}), (\ref{n=16-16}) and (\ref{n=16-17}) becomes
\begin{equation} \label{n=16-19}
-2|a_1|^2=0,
\end{equation}
and $b_2=0$ also. Therefore the $N=16$ compact Wannier function is inconsistent.

$\bm{N=19}$ \textbf{compact Wannier function}. The configuration of the compact Wannier function is shown in Supplementary Fig.~\ref{n=3 compact}e. The mirror symmetry constraints are:
\begin{subequations}
\begin{align}
&|4\rangle=\sigma_z|2\rangle\Rightarrow \left(\begin{matrix}a_4\\b_4\end{matrix}\right)=\left(\begin{matrix}a_2\\-b_2\end{matrix}\right), \label{N=19-1}\\
&\sigma_z|1\rangle=|1\rangle=\left(\begin{matrix}a_1\\0\end{matrix}\right), \label{N=19-2}\\
&\sigma_z|3\rangle=|3\rangle=\left(\begin{matrix}a_3\\0\end{matrix}\right), \label{N=19-3}\\
&\sigma_z|5\rangle=|5\rangle=\left(\begin{matrix}a_5\\0\end{matrix}\right), \label{N=19-4}\\
&\sigma_z|6\rangle=|6\rangle=\left(\begin{matrix}a_6\\0\end{matrix}\right), \label{N=19-5}\\
&\sigma_z|7\rangle=|7\rangle=\left(\begin{matrix}a_7\\0\end{matrix}\right). \label{N=19-6}
\end{align}
\end{subequations}
We only have to list the first orthogonality condition,
\begin{equation}
\langle 3|1\rangle=0\Rightarrow a_3^*a_1=0.\label{N=19-7}
\end{equation}
It is clear that either $a_1=0$ or $a_3=0$. However, both cases imply that the Wannier function becomes the $N=16$ compact Wannier function, which is inconsistent. We thus conclude that the $N=19$ compact Wannier function is inconsistent. It is clear from this case that all the Wannier function with hexagonal shape is also inconsistent.

$\bm{N=25}$ \textbf{compact Wannier function}. The configuration of the compact Wannier function is shown in Supplementary Fig.~\ref{n=3 compact}f. The mirror symmetry constraints are:
\begin{subequations}
\begin{align}
&|2\rangle=\sigma_z|1\rangle\Rightarrow \left(\begin{matrix}a_2\\b_2\end{matrix}\right)=\left(\begin{matrix}a_1\\-b_1\end{matrix}\right), \label{N=25-1}\\
&|6\rangle=\sigma_z|4\rangle\Rightarrow \left(\begin{matrix}a_6\\b_6\end{matrix}\right)=\left(\begin{matrix}a_4\\-b_4\end{matrix}\right), \label{N=25-2}\\
&\sigma_z|3\rangle=|3\rangle=\left(\begin{matrix}a_3\\0\end{matrix}\right), \label{N=25-3}\\
&\sigma_z|5\rangle=|5\rangle=\left(\begin{matrix}a_5\\0\end{matrix}\right), \label{N=25-4}\\
&\sigma_z|7\rangle=|7\rangle=\left(\begin{matrix}a_7\\0\end{matrix}\right), \label{N=25-5}\\
&\sigma_z|8\rangle=|8\rangle=\left(\begin{matrix}a_8\\0\end{matrix}\right), \label{N=25-6}\\
&\sigma_z|9\rangle=|9\rangle=\left(\begin{matrix}a_9\\0\end{matrix}\right). \label{N=25-7}
\end{align}
\end{subequations}
The orthogonality conditions require that,
\begin{subequations}
\begin{align}
&\langle 2|1\rangle=0\Rightarrow |a_1|^2-|b_1|^2=0, \label{N=25-8}\\
&\langle 5|1\rangle+\langle 2|2\rangle=0\Rightarrow a_5^*a_1+|a_1|^2+|b_1|^2=0, \label{N=25-9}\\
&\langle 6|1\rangle+\langle 2|3\rangle=0\Rightarrow a_4^*a_1-b_4^*b_1+a_1^*a_3=0, \label{N=25-10}\\
&\langle 4|1\rangle+\langle 5|3\rangle+\langle 6|2\rangle=0 \Rightarrow a_4^*a_1+b_4^*b_1+a_5^*a_3+a_4^*a_1+b_4^*b_1=0, \label{N=25-11}\\
&\langle 8|1\rangle+\langle 3|2\rangle + \langle 6|3\rangle + \langle 5|4\rangle+\langle 2|6\rangle=0 \nonumber\\
&\Rightarrow a_8^*a_1+a_3^*a_1+a_4^*a_3+a_5^*a_4+a_1^*a_4+b_1^*b_4=0, \label{N=25-12}\\
&\langle 7|1\rangle+\langle 1|2\rangle + \langle 3|3\rangle + \langle 6|4\rangle+\langle 5|5\rangle+\langle 2|7\rangle=0 \nonumber\\
&\Rightarrow a_7^*a_1+|a_3|^2+|a_4|^2-|b_4|^2+|a_5|^2+a_1^*a_7=0, \label{N=25-13}\\ 
&\langle 7|1\rangle+\langle 7|2\rangle + \langle 8|3\rangle + \langle 4|4\rangle+\langle 1|5\rangle+\langle 6|6\rangle+\langle 5|7\rangle+\langle 2|5\rangle=0 \nonumber\\
&\Rightarrow a_7^*a_1+a_7^*a_1+a_8^*a_3+|a_4|^2+|b_4|^2+a_1^*a_5+|a_4|^2+|b_4|^2+a_5^*a_7+a_1^*a_5=0, \label{N=25-14}\\
&\langle 9|1\rangle+\langle 4|2\rangle + \langle 7|3\rangle+\langle 8|4\rangle+\langle 4|5\rangle+\langle 3|6\rangle+\langle 6|7\rangle+\langle 5|8\rangle+\langle 2|8\rangle=0,\\
&\langle 8|1\rangle+\langle 8|2\rangle + \langle 9|3\rangle+\langle 7|4\rangle+\langle 3|5\rangle+\langle 7|6\rangle+\langle 8|7\rangle+\langle 4|8\rangle+\langle 1|6\rangle+\langle 3|5\rangle+\langle 6|8\rangle+\langle 5|9\rangle+\langle 2|4\rangle=0,\\
&\langle 7|1\rangle+\langle 5|2\rangle + \langle 8|3\rangle+\langle 9|4\rangle+\langle 7|5\rangle+\langle 4|6\rangle+\langle 7|7\rangle+\langle 8|8\rangle \nonumber\\
&+\langle 4|6\rangle+\langle 1|2\rangle+\langle 1|5\rangle+\langle 3|8\rangle+\langle 6|9\rangle+\langle 5|7\rangle+\langle 2|7\rangle=0, \\
&\langle 4|1\rangle+\langle 6|2\rangle + \langle 7|3\rangle+\langle 8|4\rangle+\langle 6|5\rangle+\langle 8|6\rangle+\langle 9|7\rangle+\langle 7|8\rangle \nonumber\\
&+\langle 3|6\rangle+\langle 4|5\rangle+\langle 7|8\rangle+\langle 8|9\rangle+\langle 4|7\rangle+\langle 1|3\rangle+\langle 3|4\rangle+\langle 6|7\rangle+\langle 5|8\rangle+\langle 2|3\rangle=0.
\end{align}
\end{subequations}
From Eq.~(\ref{N=25-8}) and (\ref{N=25-9}), we obtain
\begin{equation}
2|a_1|^2+a_5^*a_1=0.\label{N=25-15}
\end{equation}
If $a_1\neq 0$, then
\begin{equation}
a_5=-2a_1.\label{N=25-16}
\end{equation}
From Eq.~(\ref{N=25-11}) and (\ref{N=25-16}), it follows that
\begin{equation}
a_4^*a_1+b_4^*b_1-a_1^*a_3=0.\label{N=25-17}
\end{equation}
By combining Eq.~(\ref{N=25-10}) and (\ref{N=25-17}), we get
\begin{equation}
a_4^*a_1=0,\label{N=25-18}
\end{equation}
\begin{equation}
b_4^*b_1=a_1^*a_3,\label{N=25-19}
\end{equation}
and it follows that 
\begin{equation}
a_4=0,\label{N=25-20}
\end{equation}
\begin{equation}
|b_4|^2=|a_3|^2,\label{N=25-21}
\end{equation}
since $a_1\neq 0$ and $|a_1|^2=|b_1|^2$. From Eq.~(\ref{N=25-12}), (\ref{N=25-19}) and (\ref{N=25-20}), we obtain
\begin{equation}
a_8^*a_1+2a_3^*a_1=0,\label{N=25-22}
\end{equation}
and therefore
\begin{equation}
a_8=-2a_3.\label{N=25-23}
\end{equation}
Using Eq.~(\ref{N=25-16}), (\ref{N=25-20}) and (\ref{N=25-21}), we can rewrite Eq.~(\ref{N=25-13}) as
\begin{equation}
4|a_1|^2+a_7^*a_1+a_1^*a_7=0.\label{N=25-24}
\end{equation}
From Eq.~(\ref{N=25-14}), (\ref{N=25-16}), (\ref{N=25-20}), (\ref{N=25-21}) and (\ref{N=25-23}), it follows that
\begin{equation}
2a_7^*a_1-2a_1^*a_7=4|a_1|^2.\label{N=25-25}
\end{equation}
It is clear that the only solution is $a_1=0$ which is in contradiction to our assumption. Therefore $a_1=0$ in the first place, but it implies that $a_1=b_1=a_2=b_2=0$. The Wannier function then becomes the $N=19$ compact Wannier function which is inconsistent. We thus conclude that the $N=25$ compact Wannier function is inconsistent.

\textbf{Type II CDI\bm{$_3$}.} The mirror reflection planes pass through the center of the equilateral triangle, see Supplementary Fig.~\ref{n=3-2 compact}a. The three mirror symmetries can be denoted as $\mathcal{M}_x,C_3\mathcal{M}_xC_3^{-1},C_3^2\mathcal{M}_xC_3^{-2}$. The basis orbitals consist of a $s$ orbital and a $f_{x(x^2-3y^2)}$ orbital.

\begin{figure}
\begin{center}
\includegraphics[width=5in]{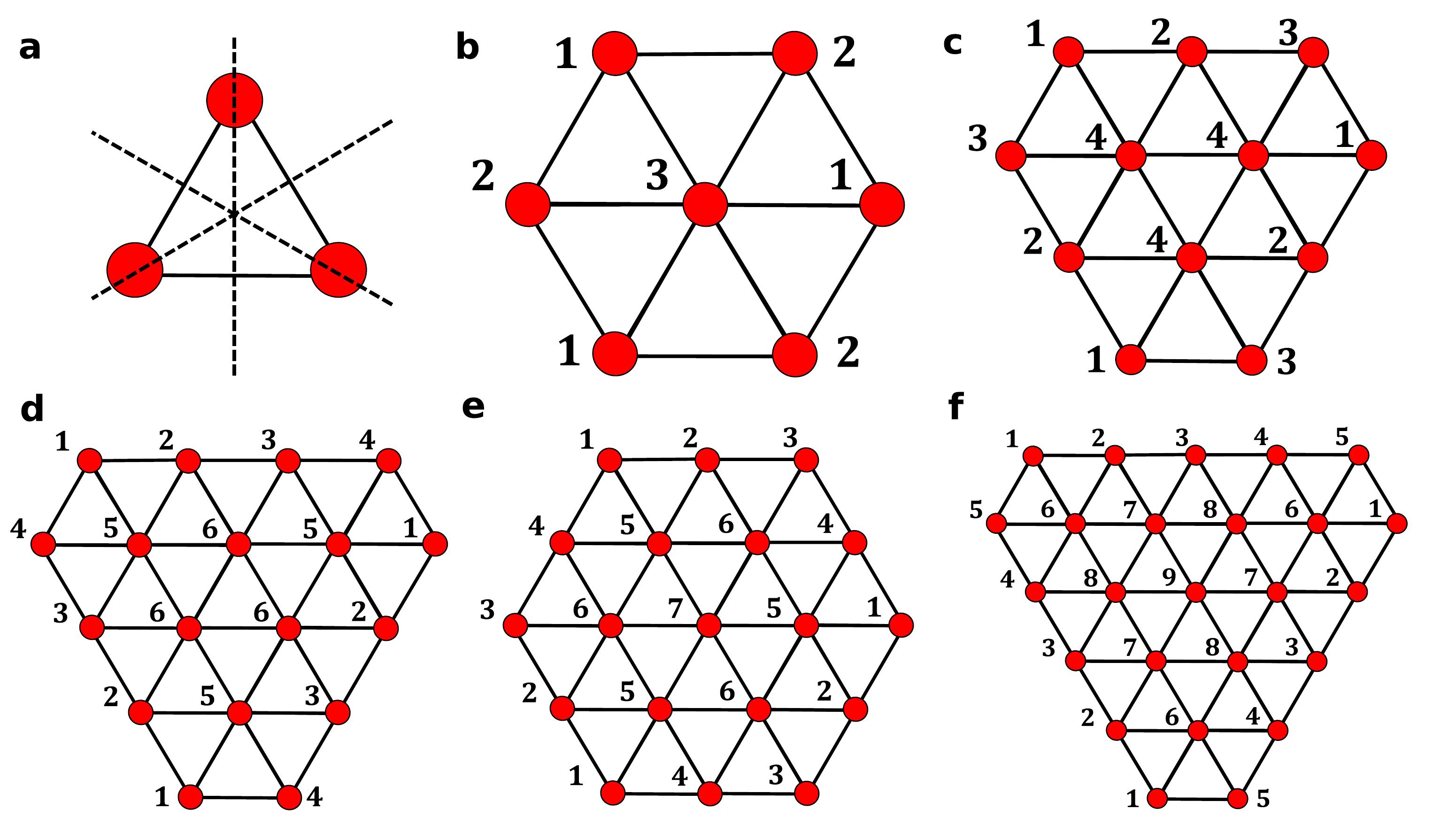}
\caption[]{Supplementary Figure 6. \textbf{Compact Wannier functions for the Type II $\bm{n=3}$ Chern dartboard insulator.} $\textbf{\textsf{a}}$ An equilateral triangular shape of the Wannier function. The dashed lines show the mirror reflection planes. $\textbf{\textsf{b}}$ $N=7$ compact Wannier function. $\textbf{\textsf{c}}$ $N=12$ compact Wannier function. $\textbf{\textsf{d}}$ $N=18$ compact Wannier function. $\textbf{\textsf{e}}$ $N=19$ compact Wannier function. $\textbf{\textsf{f}}$ $N=25$ compact Wannier function.}
\label{n=3-2 compact}
\end{center}
\end{figure}

$\bm{N=3}$ \textbf{compact Wannier function}. The configuration of the compact Wannier function is shown in Supplementary Fig.~\ref{n=3-2 compact}a. Since the mirror reflection planes pass through each atom that we label as $1$, the atomic lattice states should be the eigenstates of the mirror symmetry representation $\sigma_z$ with eigenvalue $1$. That is,
\begin{equation}
\sigma_z|1\rangle=|1\rangle=\left(\begin{matrix}a_1\\0\end{matrix}\right).
\end{equation}
However, the orthogonality condition requires that  
\begin{equation}
\langle 1|1\rangle=1/\sqrt{3}=0,
\end{equation}
which is impossible. We note that this contradiction arises as long as the shape of the compact Wannier function is an equilateral triangle (or star shape) regardless of the total number of the occupied atomic lattice sites inside. Therefore, in the following, we have automatically excluded the compact Wannier functions with this kind of shape.

$\bm{N=7}$ \textbf{compact Wannier function}. The configuration of the compact Wannier function is shown in Supplementary Fig.~\ref{n=3-2 compact}b. First we list the mirror symmetry constraints:
\begin{subequations}
\begin{align}
&|2\rangle=\sigma_z|1\rangle\Rightarrow \left(\begin{matrix}a_2\\b_2\end{matrix}\right)=\left(\begin{matrix}a_1\\-b_1\end{matrix}\right), \label{n=7-1}\\
&\sigma_z|3\rangle=|3\rangle=\left(\begin{matrix}a_3\\0\end{matrix}\right).\label{n=7-2}
\end{align}
\end{subequations}
The orthogonality conditions require that,
\begin{subequations}
\begin{align}
&\langle 2|1\rangle=0\Rightarrow |a_1|^2-|b_1|^2=0, \label{n=7-3}\\
&\langle 1|1\rangle+\langle 2|2\rangle=0\Rightarrow |a_1|^2+|b_1|^2+|a_1|^2+|b_1|^2=0, \label{n=7-4}\\
&\langle 3|1\rangle+\langle 1|2\rangle + \langle 1|2\rangle + \langle 2|3\rangle=0 \label{n=7-5}.
\end{align}
\end{subequations}
From Eq.~(\ref{n=7-1}) and (\ref{n=7-4}), we can conclude that,
\begin{equation}
a_1=b_1=a_2=b_2=0,
\end{equation}
and the $N=7$ compact Wannier function is inconsistent.

$\bm{N=12}$ \textbf{compact Wannier function}. The configuration of the compact Wannier function is shown in Supplementary Fig.~\ref{n=3-2 compact}c. The mirror symmetry constraints are:
\begin{subequations}
\begin{align}
&|3\rangle=\sigma_z|1\rangle\Rightarrow \left(\begin{matrix}a_3\\b_3\end{matrix}\right)=\left(\begin{matrix}a_1\\-b_1\end{matrix}\right), \label{n=12-1}\\
&\sigma_z|2\rangle=|2\rangle=\left(\begin{matrix}a_2\\0\end{matrix}\right), \label{n=12-2}\\
&\sigma_z|4\rangle=|4\rangle=\left(\begin{matrix}a_4\\0\end{matrix}\right). \label{n=12-3}
\end{align}
\end{subequations}
The orthogonality conditions require that,
\begin{subequations}
\begin{align}
&\langle 3|1\rangle=0\Rightarrow |a_1|^2-|b_1|^2=0, \label{n=12-4}\\
&\langle 1|1\rangle+\langle 3|2\rangle=0\Rightarrow |a_1|^2+|b_1|^2+a_1^*a_2=0, \label{n=12-5}\\
&\langle 4|1\rangle+\langle 2|2\rangle+\langle 1|3\rangle + \langle 3|4\rangle=0\Rightarrow a_4^*a_1+|a_2|^2+a_1^*a_4=0, \label{n=12-6}\\
&\langle 2|1\rangle+\langle 4|2\rangle + \langle 2|3\rangle + \langle 1|4\rangle+\langle 3|4\rangle=0\Rightarrow a_2^*a_1+a_4^*a_2+a_2^*a_1+a_1^*a_4+a_1^*a_4=0, \label{n=12-7}\\
&\langle 4|1\rangle+\langle 4|2\rangle + \langle 1|3\rangle + \langle 2|3\rangle+\langle 4|4\rangle+\langle 2|4\rangle+\langle 1|2\rangle+\langle 3|4\rangle=0. \label{n=12-8}
\end{align}
\end{subequations}
From Eq.~(\ref{n=12-4}) and (\ref{n=12-5}), we can obtain
\begin{equation} \label{n=12-9}
a_1^*(2a_1+a_2)=0.
\end{equation}
If $a_1\neq0$, this implies that
\begin{equation} \label{n=12-10}
a_2=-2a_1.
\end{equation}
By plugging Eq.~(\ref{n=12-10}) into (\ref{n=12-6}) and (\ref{n=12-7}), it follows that
\begin{equation} \label{n=12-11}
\begin{split}
4|a_1|^2+a_1^*a_4+a_4^*a_1&=0,\\
-4|a_1|^2+2a_1^*a_4-2a_4^*a_1&=0.
\end{split}
\end{equation}
It is clear that the only solution is $a_1=0$ which is in contradiction to our assumption. Therefore $a_1=0$ in the first place. But this implies that $b_1=a_3=b_3=0$, and the shape of the Wannier function becomes an equilateral triangle. Therefore the $N=12$ compact Wannier function is inconsistent.

$\bm{N=18}$ \textbf{compact Wannier function}. The configuration of the compact Wannier function is shown in Supplementary Fig.~\ref{n=3-2 compact}d. The mirror symmetry constraints are:
\begin{subequations}
\begin{align}
&|4\rangle=\sigma_z|1\rangle\Rightarrow \left(\begin{matrix}a_4\\b_4\end{matrix}\right)=\left(\begin{matrix}a_1\\-b_1\end{matrix}\right), \label{n=18-1}\\
&|3\rangle=\sigma_z|2\rangle\Rightarrow \left(\begin{matrix}a_3\\b_3\end{matrix}\right)=\left(\begin{matrix}a_2\\-b_2\end{matrix}\right), \label{n=18-2}\\
&\sigma_z|5\rangle=|5\rangle=\left(\begin{matrix}a_5\\0\end{matrix}\right), \label{n=18-3}\\
&\sigma_z|6\rangle=|6\rangle=\left(\begin{matrix}a_6\\0\end{matrix}\right). \label{n=18-4}
\end{align}
\end{subequations}
The orthogonality conditions require that,
\begin{subequations}
\begin{align}
&\langle 4|1\rangle=0\Rightarrow |a_1|^2-|b_1|^2=0, \label{n=18-5}\\
&\langle 1|1\rangle+\langle 4|2\rangle=0\Rightarrow |a_1|^2+|b_1|^2+a_1^*a_2-b_1^*b_2=0, \label{n=18-6}\\
&\langle 1|2\rangle+\langle 4|3\rangle=0\Rightarrow a_1^*a_2+b_1^*b_2+a_1^*a_2+b_1^*b_2=0, \label{n=18-7}\\
&\langle 5|1\rangle+\langle 3|2\rangle+\langle 1|4\rangle + \langle 4|5\rangle=0\Rightarrow a_5^*a_1+|a_2|^2-|b_2|^2+a_1^*a_5=0, \label{n=18-8}\\
&\langle 2|1\rangle+\langle 5|2\rangle + \langle 3|3\rangle + \langle 1|5\rangle+\langle 4|6\rangle=0 \Rightarrow a_2^*a_1+b_2^*b_1+a_5^*a_2+|a_2|^2+|b_2|^2+a_1^*a_5+a_1^*a_6=0, \label{n=18-9}\\
&\langle 6|1\rangle+\langle 6|2\rangle + \langle 2|3\rangle + \langle 2|4\rangle+\langle 5|5\rangle+\langle 3|6\rangle+\langle 1|3\rangle+\langle 4|6\rangle=0 \nonumber\\
&\Rightarrow a_6^*a_1+a_6^*a_2+|a_2|^2-|b_2|^2+a_2^*a_1-b_2^*b_1+|a_5|^2+a_2^*a_6+a_1^*a_2-b_1^*b_2+a_1^*a_6=0, \label{n=18-10}\\ 
&\langle 3|1\rangle+\langle 6|2\rangle + \langle 6|3\rangle + \langle 2|4\rangle+\langle 2|5\rangle+\langle 5|6\rangle+\langle 3|5\rangle+\langle 1|6\rangle+\langle 4|6\rangle=0 \nonumber\\
&\Rightarrow a_2^*a_1-b_2^*b_1+a_6^*a_2+a_6^*a_2+a_2^*a_1-b_2^*b_1+a_2^*a_5+a_5^*a_6+a_2^*a_5+a_1^*a_6+a_1^*a_6=0, \label{n=18-11}\\
&\langle 5|1\rangle+\langle 6|2\rangle + \langle 5|3\rangle + \langle 4|1\rangle+\langle 3|4\rangle+\langle 6|5\rangle+\langle 6|6\rangle+\langle 2|5\rangle+\langle 2|3\rangle+\langle 5|6\rangle+\langle 3|6\rangle+\langle 1|2\rangle+\langle 4|5\rangle=0. \label{n=18-12}
\end{align}
\end{subequations}
From Eq.~(\ref{n=18-5}), (\ref{n=18-6}) and (\ref{n=18-7}), we can obtain
\begin{equation} \label{n=18-13}
a_1^*(a_1+a_2)=0.
\end{equation}
If $a_1\neq0$, it implies that
\begin{equation} \label{n=18-14}
a_2=-a_1.
\end{equation}
By plugging Eq.~(\ref{n=18-14}) into (\ref{n=18-6}) or (\ref{n=18-7}), it follows that
\begin{equation} \label{n=18-15}
b_1^*(b_1-b_2)=0.
\end{equation}
Since $a_1\neq 0$ and Eq.~(\ref{n=18-5}), it implies that $b_1\neq 0$ and 
\begin{equation} \label{n=18-16}
b_2=b_1.
\end{equation}
Using the results of Eq.~(\ref{n=18-5}), (\ref{n=18-14}) and (\ref{n=18-16}), we can get from Eq.~(\ref{n=18-8}) and (\ref{n=18-9})
\begin{equation} \label{n=18-17}
a_5^*a_1+a_1^*a_5=0.
\end{equation}
\begin{equation} \label{n=18-18}
-a_5^*a_1+a_1^*a_5+2|a_1|^2+a_1^*a_6=0.
\end{equation}
Combining the above equations, it follows that
\begin{equation} \label{n=18-19}
2a_1+2a_5+a_6=0.
\end{equation}
By plugging Eq.~(\ref{n=18-5}), (\ref{n=18-14}) and (\ref{n=18-16}) into Eq.~(\ref{n=18-10}), we obtain
\begin{equation} \label{n=18-20}
|a_5|^2=4|a_1|^2.
\end{equation}
Remember we still have two orthogonality conditions to check. Luckily, the first one Eq.~(\ref{n=18-11}) is enough, which gives
\begin{equation} \label{n=18-21}
-4|a_1|^2-2a_6^*a_1-2a_1^*a_5+a_5^*a_6+2a_1^*a_6=0.
\end{equation}
Using Eq.~(\ref{n=18-17}), (\ref{n=18-19}) and (\ref{n=18-20}), we can rewrite this equation as
\begin{equation} \label{n=18-22}
-12|a_1|^2-8a_1^*a_5=0.
\end{equation}
It is clear that the only solution to Eq.~(\ref{n=18-20}) and Eq.~(\ref{n=18-22}) is $a_1=0$ which is in contradiction to our assumption. Therefore $a_1=0$ in the first place. But this implies that $b_1=a_4=b_4=0$, and the shape of the Wannier function becomes the $N=12$ compact Wannier function. Therefore the $N=18$ compact Wannier function is inconsistent.

$\bm{N=19}$ \textbf{compact Wannier function}. The configuration of the compact Wannier function is shown in Supplementary Fig.~\ref{n=3-2 compact}e. The mirror symmetry constraints are:
\begin{subequations}
\begin{align}
&|3\rangle=\sigma_z|1\rangle\Rightarrow \left(\begin{matrix}a_3\\b_3\end{matrix}\right)=\left(\begin{matrix}a_1\\-b_1\end{matrix}\right), \label{n=19-1}\\
&|6\rangle=\sigma_z|5\rangle\Rightarrow \left(\begin{matrix}a_6\\b_6\end{matrix}\right)=\left(\begin{matrix}a_5\\-b_5\end{matrix}\right), \label{n=19-2}\\
&\sigma_z|2\rangle=|2\rangle=\left(\begin{matrix}a_2\\0\end{matrix}\right), \label{n=19-3}\\
&\sigma_z|4\rangle=|4\rangle=\left(\begin{matrix}a_4\\0\end{matrix}\right), \label{n=19-4}\\
&\sigma_z|7\rangle=|7\rangle=\left(\begin{matrix}a_7\\0\end{matrix}\right). \label{n=19-5}
\end{align}
\end{subequations}
The orthogonality conditions require that,
\begin{subequations}
\begin{align}
&\langle 3|1\rangle=0\Rightarrow |a_1|^2-|b_1|^2=0, \label{n=19-6}\\
&\langle 4|1\rangle+\langle 3|2\rangle=0\Rightarrow a_4^*a_1+a_1^*a_2=0, \label{n=19-7}\\
&\langle 1|1\rangle+\langle 4|2\rangle+\langle 3|3\rangle=0 \Rightarrow|a_1|^2+|b_1|^2+a_4^*a_2+|a_1|^2+|b_1|^2=0, \label{n=19-8}\\
&\langle 6|1\rangle+\langle 2|2\rangle+\langle 4|4\rangle + \langle 3|5\rangle=0 \Rightarrow a_5^*a_1-b_5^*b_1+|a_2|^2+|a_4|^2+a_1^*a_5-b_1^*b_5=0, \label{n=19-9}\\
&\langle 5|1\rangle+\langle 6|2\rangle + \langle 2|3\rangle + \langle 1|4\rangle+\langle 4|5\rangle+\langle 3|6\rangle=0 \nonumber\\
&\Rightarrow a_5^*a_1+b_5^*b_1+a_5^*a_2+a_2^*a_1+a_1^*a_4+a_4^*a_5+a_1^*a_5+b_1^*b_5=0, \label{n=19-10}\\
&\langle 7|1\rangle+\langle 5|2\rangle + \langle 1|3\rangle + \langle 5|4\rangle+\langle 6|5\rangle+\langle 2|6\rangle+\langle 1|3\rangle+\langle 4|6\rangle+\langle 3|7\rangle=0 \nonumber\\
&\Rightarrow a_7^*a_1+a_5^*a_2+a_5^*a_4+|a_5|^2-|b_5|^2+a_2^*a_5+a_4^*a_5+a_1^*a_7=0, \label{n=19-11}\\ 
&\langle 6|1\rangle+\langle 7|2\rangle + \langle 5|3\rangle + \langle 2|4\rangle+\langle 5|5\rangle+\langle 6|6\rangle+\langle 2|4\rangle+\langle 1|6\rangle+\langle 4|7\rangle+\langle 3|5\rangle=0 \nonumber\\
&\Rightarrow a_5^*a_1-b_5^*b_1+a_7^*a_2+a_5^*a_1-b_5^*b_1+2a_2^*a_4+2|a_5|^2+2|b_5|^2+a_1^*a_5-b_1^*b_5+a_4^*a_7+a_1^*a_5-b_1^*b_5=0, \label{n=19-12}\\
&\langle 5|1\rangle+\langle 6|2\rangle + \langle 4|3\rangle+\langle 6|4\rangle+\langle 7|5\rangle+\langle 5|6\rangle+\langle 1|4\rangle+\langle 2|3\rangle+\langle 5|6\rangle+\langle 6|7\rangle+\langle 2|5\rangle+\langle 1|2\rangle+\langle 4|5\rangle+\langle 3|6\rangle=0 \nonumber\\
&\Rightarrow a_5^*a_1+b_5^*b_1+a_5^*a_2+a_4^*a_1+a_5^*a_4+a_7^*a_5+2|a_5|^2-2|b_5|^2+a_1^*a_4+a_2^*a_1+a_5^*a_7+a_2^*a_5+a_1^*a_2+a_4^*a_5 \nonumber\\
&+a_1^*a_5+b_1^*b_5=0. \label{n=19-13}
\end{align}
\end{subequations}
From Eq.~(\ref{n=19-6}), (\ref{n=19-7}) and (\ref{n=19-8}), we can obtain
\begin{equation} \label{n=19-14}
4|a_1|^2a_1-a_1^*a_2^2=0.
\end{equation}
If $a_1\neq 0$,
\begin{equation} \label{n=19-15}
a_2=\pm 2a_1.
\end{equation}
By plugging it into Eq.~(\ref{n=19-7}), we get
\begin{equation} \label{n=19-16}
a_4=\mp 2a_1.
\end{equation}
From Eq.~(\ref{n=19-15}) and (\ref{n=19-16}), we can simplify Eq.~(\ref{n=19-9}) and (\ref{n=19-10}) into
\begin{equation} \label{n=19-17}
8|a_1|^2+a_5^*a_1+a_1^*a_5-b_5^*b_1-b_1^*b_5=0,
\end{equation}
\begin{equation} \label{n=19-18}
(1\pm 2)a_5^*a_1+(1\mp 2)a_1^*a_5+b_5^*b_1+b_1^*b_5=0.
\end{equation}
Combining these two equations, it follows that
\begin{equation} \label{n=19-19}
8|a_1|^2+(2\pm 2)a_5^*a_1+(2\mp 2)a_1^*a_5=0,
\end{equation}
and
\begin{equation} \label{n=19-20}
a_5=-2a_1.
\end{equation}
Then Eq.~(\ref{n=19-17}) can be rewritten as
\begin{equation} \label{n=19-21}
b_5^*b_1+b_1^*b_5=4|a_1|^2.
\end{equation}
Using Eq.~(\ref{n=19-15}), (\ref{n=19-16}) and (\ref{n=19-20}), Eq.~(\ref{n=19-11}) and (\ref{n=19-12}) become
\begin{equation} \label{n=19-22}
4|a_1|^2-|b_5|^2+a_7^*a_1+a_1^*a_7=0,
\end{equation}
\begin{equation} \label{n=19-23}
-8|a_1|^2+2|b_5|^2-2b_5^*b_1-2b_1^*b_5\pm 2a_7^*a_1\mp 2a_1^*a_7=0.
\end{equation}
From Eq.~(\ref{n=19-21}), (\ref{n=19-22}) and (\ref{n=19-23}), it follows that
\begin{equation} \label{n=19-24}
-8|a_1|^2+(2\pm 2)a_7^*a_1+(2\mp 2)a_1^*a_7=0,
\end{equation}
and 
\begin{equation} \label{n=19-25}
a_7=2a_1.
\end{equation}
Plugging it into Eq.~(\ref{n=19-22}), we obtain
\begin{equation} \label{n=19-26}
|b_5|^2=8|a_1|^2.
\end{equation}
We have still one orthogonality condition Eq.~(\ref{n=19-13}) to check. Using Eq.~(\ref{n=19-15}), (\ref{n=19-16}), (\ref{n=19-20}), (\ref{n=19-25}) and (\ref{n=19-26}), we can simplify it into
\begin{equation} \label{n=19-27}
b_5^*b_1+b_1^*b_5=20|a_1|^2.
\end{equation}
From Eq.~(\ref{n=19-21}) and (\ref{n=19-27}), it is clear that the only solution is $a_1=0$ which is in contradiction to our assumption. Therefore $a_1=0$ in the first place. But it implies that $b_1=a_3=b_3=a_4=a_2=0$, and the shape of the Wannier function becomes the $N=7$ compact Wannier function. Therefore the $N=19$ compact Wannier function is inconsistent.

$\bm{N=25}$ \textbf{compact Wannier function}. The configuration of the compact Wannier function is shown in Supplementary Fig.~\ref{n=3-2 compact}f. The mirror symmetry constraints are:
\begin{subequations}
\begin{align}
&|5\rangle=\sigma_z|1\rangle\Rightarrow \left(\begin{matrix}a_5\\b_5\end{matrix}\right)=\left(\begin{matrix}a_1\\-b_1\end{matrix}\right), \label{n=25-1}\\
&|4\rangle=\sigma_z|2\rangle\Rightarrow \left(\begin{matrix}a_4\\b_4\end{matrix}\right)=\left(\begin{matrix}a_2\\-b_2\end{matrix}\right), \label{n=25-2}\\
&|8\rangle=\sigma_z|7\rangle\Rightarrow \left(\begin{matrix}a_8\\b_8\end{matrix}\right)=\left(\begin{matrix}a_7\\-b_7\end{matrix}\right), \label{n=25-3}\\
&\sigma_z|3\rangle=|3\rangle=\left(\begin{matrix}a_3\\0\end{matrix}\right), \label{n=25-4}\\
&\sigma_z|6\rangle=|6\rangle=\left(\begin{matrix}a_6\\0\end{matrix}\right), \label{n=25-5}\\
&\sigma_z|9\rangle=|9\rangle=\left(\begin{matrix}a_9\\0\end{matrix}\right). \label{n=25-6}
\end{align}
\end{subequations}
The orthogonality conditions require that,
\begin{subequations}
\begin{align}
&\langle 5|1\rangle=0\Rightarrow |a_1|^2-|b_1|^2=0, \label{n=25-7}\\
&\langle 1|1\rangle+\langle 5|2\rangle=0\Rightarrow |a_1|^2+|b_1|^2+a_1^*a_2-b_1^*b_2=0, \label{n=25-8}\\
&\langle 1|2\rangle+\langle 5|3\rangle=0\Rightarrow a_1^*a_2+b_1^*b_2+a_1^*a_3=0, \label{n=25-9}\\
&\langle 6|1\rangle+\langle 4|2\rangle+\langle 1|5\rangle + \langle 5|6\rangle=0 \Rightarrow a_6^*a_1+|a_2|^2-|b_2|^2+a_1^*a_6=0, \label{n=25-10}\\
&\langle 2|1\rangle+\langle 6|2\rangle + \langle 4|3\rangle + \langle 1|6\rangle+\langle 5|7\rangle=0 \Rightarrow a_2^*a_1+b_2^*b_1+a_6^*a_2+a_2^*a_3+a_1^*a_6+a_1^*a_7-b_1^*b_7=0, \label{n=25-11}\\
&\langle 2|2\rangle+\langle 6|3\rangle + \langle 4|4\rangle + \langle 1|7\rangle+\langle 5|8\rangle=0 \Rightarrow 2|a_2|^2+2|b_2|^2+a_6^*a_3+2a_1^*a_7+2b_1^*b_7=0, \label{n=25-12}\\
&\langle 7|1\rangle+\langle 8|2\rangle + \langle 3|3\rangle + \langle 2|5\rangle+\langle 6|6\rangle+\langle 4|7\rangle+\langle 1|4\rangle+\langle 5|8\rangle=0 \nonumber\\
&\Rightarrow a_7^*a_1+b_7^*b_1+a_7^*a_2-b_7^*b_2+|a_3|^2+a_2^*a_1-b_2^*b_1+|a_6|^2+a_2^*a_7-b_2^*b_7+a_1^*a_2-b_1^*b_2+a_1^*a_7+b_1^*b_7=0, \label{n=25-13}\\
&\langle 3|1\rangle+\langle 7|2\rangle + \langle 8|3\rangle+\langle 3|4\rangle+\langle 2|6\rangle+\langle 6|7\rangle+\langle 4|8\rangle+\langle 1|8\rangle+\langle 5|9\rangle=0 \nonumber\\
&\Rightarrow a_3^*a_1+a_7^*a_2+b_7^*b_2+a_7^*a_3+a_3^*a_2+a_2^*a_6+a_6^*a_7+a_2^*a_7+b_2^*b_7+a_1^*a_7-b_1^*b_7+a_1^*a_9=0, \label{n=25-14}\\
&\langle 8|1\rangle+\langle 9|2\rangle + \langle 7|3\rangle+\langle 2|4\rangle+\langle 3|5\rangle+\langle 7|6\rangle+\langle 8|7\rangle+\langle 3|8\rangle+\langle 2|4\rangle+\langle 6|8\rangle+\langle 4|9\rangle+\langle 1|3\rangle+\langle 5|7\rangle=0 \nonumber\\
&\Rightarrow a_7^*a_1-b_7^*b_1+a_9^*a_2+a_7^*a_3+2|a_2|^2-2|b_2|^2+a_3^*a_1+a_7^*a_6 \nonumber\\
&+|a_7|^2-|b_7|^2+a_3^*a_7+a_6^*a_7+a_2^*a_9+a_1^*a_3+a_1^*a_7-b_1^*b_7=0, \label{n=25-15}\\
&\langle 4|1\rangle+\langle 8|2\rangle + \langle 9|3\rangle+\langle 7|4\rangle+\langle 2|5\rangle+\langle 3|6\rangle+\langle 7|7\rangle+\langle 8|8\rangle+\langle 3|6\rangle+\langle 2|8\rangle \nonumber\\
&+\langle 6|9\rangle+\langle 4|7\rangle+\langle 1|7\rangle+\langle 5|8\rangle=0, \nonumber\\
&\Rightarrow 2a_2^*a_1-2b_2^*b_1+2a_7^*a_2-2b_7^*b_2+a_9^*a_3+2a_3^*a_6\nonumber\\
&+2|a_7|^2+2|b_7|^2+2a_2^*a_7-2b_2^*b_7+a_6^*a_9+2a_1^*a_7+2b_1^*b_7=0, \label{n=25-16}\\
&\langle 6|1\rangle+\langle 7|2\rangle + \langle 8|3\rangle+\langle 6|4\rangle+\langle 1|5\rangle+\langle 4|5\rangle+\langle 8|6\rangle+\langle 9|7\rangle+\langle 7|8\rangle+\langle 2|6\rangle\\
&+\langle 3|4\rangle+\langle 7|8\rangle+\langle 8|9\rangle+\langle 3|7\rangle+\langle 2|3\rangle+\langle 6|7\rangle+\langle 4|8\rangle+\langle 1|2\rangle+\langle 5|6\rangle=0. \nonumber \label{n=25-17}
\end{align}
\end{subequations}
From Eq.~(\ref{n=25-7}), (\ref{n=25-8}) and (\ref{n=25-9}), we can obtain
\begin{equation} \label{n=25-18}
2|a_1|^2+2a_1^*a_2+a_1^*a_3=0.
\end{equation}
If $a_1\neq 0$,
\begin{equation} \label{n=25-19}
a_3=-2a_1-2a_2.
\end{equation}
We can solve for $b_2$ from Eq.~(\ref{n=25-7}) and (\ref{n=25-8}),
\begin{equation} \label{n=25-20}
b_2=\frac{2|a_1|^2+a_1^*a_2}{b_1^*}.
\end{equation}
Notice that $b_1\neq 0$ since $a_1\neq 0$. From Eq.~(\ref{n=25-8}), (\ref{n=25-10}), (\ref{n=25-11}) and (\ref{n=25-12}), we get
\begin{equation} \label{n=25-21}
|a_1|^2+a_1^*a_6+a_1^*a_7=0,
\end{equation}
and therefore
\begin{equation} \label{n=25-22}
a_6=-a_1-a_7.
\end{equation}
By using Eq.~(\ref{n=25-20}) and (\ref{n=25-22}), Eq.~(\ref{n=25-10}) can be rewritten as
\begin{equation} \label{n=25-23}
6|a_1|^2+2a_1^*a_2+2a_2^*a_2+a_1^*a_7+a_7^*a_1=0.
\end{equation}
One can observe that the quantity 
\begin{equation} \label{n=25-24}
3|a_1|^2+2a_1^*a_2+a_1^*a_7\equiv i_7
\end{equation}
must be an imaginary number. Now we can solve for $a_7$,
\begin{equation} \label{n=25-25}
a_7=-3a_1-2a_2-\frac{i_7}{a_1^*}.
\end{equation}
Using Eq.~(\ref{n=25-19}), (\ref{n=25-20}), (\ref{n=25-22}) and (\ref{n=25-25}), we can solve for $b_7$ from Eq.~(\ref{n=25-10}),
\begin{equation} \label{n=25-26}
b_7=\frac{|a_1|^2+2a_1^*a_2-i_7a_2/a_1}{b_1^*}.
\end{equation}
Plugging Eq.~(\ref{n=25-19}), (\ref{n=25-20}), (\ref{n=25-22}), (\ref{n=25-25}) and (\ref{n=25-26}) into Eq.~(\ref{n=25-13}), it follows that
\begin{equation} \label{n=25-27}
\frac{|i_7|^2}{|a_1|^2}=4|a_1|^2.
\end{equation}
and therefore
\begin{equation} \label{n=25-28}
i_7=\pm 2|a_1|^2i.
\end{equation}
We can obtain the expressions for $a_6$, $a_7$ and $b_7$,
\begin{equation} \label{n=25-29}
a_6=(2\pm 2i)a_1+2a_2,
\end{equation}
\begin{equation} \label{n=25-30}
a_7=-(3\pm 2i)a_1-2a_2,
\end{equation}
\begin{equation} \label{n=25-31}
b_7=\frac{|a_1|^2+(2\mp 2i)a_1^*a_2}{b_1^*}.
\end{equation}
Eq.~(\ref{n=25-14}) can thus be simplified into
\begin{equation} \label{n=25-32}
(-6\mp 4i)|a_1|^2-2a_1^*a_2+a_1^*a_9=0,
\end{equation}
and consequently
\begin{equation} \label{n=25-33}
a_9=(6\pm 4i)a_1+2a_2.
\end{equation}
We have still three orthogonality conditions to check. The first one Eq.~(\ref{n=25-15}) after plugging all the solved variables gives
\begin{equation} \label{n=25-34}
-8|a_1|^2+2|a_2|^2+4a_1^*a_2+4a_2^*a_1=0.
\end{equation}
The second one Eq.~(\ref{n=25-16}) gives
\begin{equation} \label{n=25-35}
(12\mp 8i)|a_1|^2+(-12\mp 4i)a_1^*a_2+(4\pm 4i)a_2^*a_1=0.
\end{equation}
Now we can solve for $a_2$,
\begin{equation} \label{n=25-36}
a_2=\left(1\mp\frac{1}{2}i\right)a_1.
\end{equation}
By plugging it into Eq.~(\ref{n=25-34}), we obtain
\begin{equation} \label{n=25-38}
|a_2|^2=0.
\end{equation}
Therefore $a_1=a_2=0$ which is in contradiction to our assumption, and thus $a_1=0$ in the first place. But it implies that $b_1=a_5=b_5=0$, and the shape of the Wannier function becomes the $N=19$ compact Wannier function. We can conclude accordingly that the $N=25$ compact Wannier function is inconsistent.